\documentstyle[clbiba,numreferences]{kapproc}

\font \math=msbm10 scaled \magstep 0

\newcommand{\mmath}[1]{{\mbox{\math #1}}}

\newcommand{\Co}{\mmath{C}}

\newcommand{\Ga}{\mmath{Z}}

\newcommand{\MX}{\mmath{X}}

\newcommand{\RE}{\mmath{R}}
\newcommand{\ot}{\frac{1}{2}}

\newcommand{\be}{\begin{equation}}
\newcommand{\ee}{\end{equation}}
\newcommand{\bea}{\begin{eqnarray}}
\newcommand{\eea}{\end{eqnarray}}
\newcommand{\gsim}{\stackrel{>}{\raisebox{-1.5pt}{$\sim$}}} 
\newcommand{\lsim}{\stackrel{<}{\raisebox{-1.5pt}{$\sim$}}} 
\newcommand{\eq}[1]{(\ref{#1})}
\newcommand{\ZETA}{\mbox{\boldmath $\zeta$}}
\newcommand{\QA}{\;, \quad }
\renewcommand{\theequation}{\thesection.\arabic{equation}}
\newcommand{\dising}{\delta_{\mbox{\scriptsize Ising}}}

\newcommand{\vecu}{{\bf u}}
\newcommand{\vecv}{{\bf v}}

\newcommand{\FIG}[6]{
\begin{figure}[ht]
\hfil \hbox to #4{\vbox to #5{\vfil
 } }
\hfil
\caption{\sloppy #3 \label{#1}}
\end{figure}}

\begin{opening}
\title{Computer Stochastics in Scalar Quantum Field
Theory\thanks{to appear in {\em Stochastic Analysis and Applications
in Physics}, Proc. of the NATO ASI in Funchal, Madeira, Aug. 1993,
ed. L. Streit (Kluwer Acad. Publishers, Dordrecht: 1994)}}
\author{C.B. Lang}
\institute{Inst. f. theoret. Physik\\
Universit\"at Graz \\Universit\"atsplatz 5\\A-8010 Graz,
AUSTRIA}
\end{opening}

\begin{document}

\begin{abstract}
This is a series of lectures on Monte Carlo results on the
non-perturbative, lattice formulation approach to quantum field
theory.  Emphasis is put on 4D scalar quantum field theory. I discuss
real space renormalization group, fixed point properties and
logarithmic corrections, partition function zeroes, the triviality
bound on the Higgs mass, finite size effects, Goldstone bosons and
chiral perturbation theory, and the determination of scattering phase
shifts for some scalar models.

\keywords Scalar quantum field theory, $\Phi^4$ model, Monte Carlo
methods

\end{abstract}

\section{Lattice Quantum Field Theory}
\setcounter{equation}{0}

In this series of lectures I attempt to cover some numerical
applications to stochastic analysis in relativistic quantum field
theory (QFT). Following a more general introduction I then focus on the
non-perturbative quantization of {\em scalar\/} QFT. This emphasis is
dictated by the limited time and place available as well as by the
progress in the subject. Scalar QFT meanwhile is the most advanced of
the interesting models, both in theoretical understanding as well as in
numerical stochastic calculations.

If you are interested in particular topics of lattice field (including
gauge field) theory, a good starting point is to look up the
proceedings of the annual workshops in lattice field theory. For the
recent years they have been published in Nucl. Phys. B (Proc. Suppl.);
up to now there are volumes 9(1989), 17(1990), 20(1991), 26(1992),
30(1993). In every one of these volumes there are also reviews,
although mainly targeted at the devotees of the field.

There are only few books giving surveys. Almost all important results
of the first, hot days of Monte Carlo computations are given in the
reprint collection \cite{Re83} and the monograph \cite{Cr83}.  The more
rigorously oriented person should maybe start with \cite{Se82}.
Recently two volumes on the field have been published, one is the
monograph \cite{Ro92} and the other contains a collection of review
article written by experts on various subtopics \cite{Cr92}.

In the first part I attempt an overview on lattice field theory and the
related computational issues. The role of the $\Phi^4$ model in
particle physics is discussed and the lattice formulation for that
model is introduced. In the second chapter some rigorous statements on
the model are reviewed and then lattice calculation results concerning
the fixed point structure and the issue of triviality are presented. In
the subsequent chapter further results on the so-called triviality
bound are discussed; also the particle spectrum and in particular the
finite size effects produced by the massless Goldstone states are
reviewed. In the last section I present recent results that use the
scalar model as a testing ground for new methods to determine
scattering phase shifts non-perturbatively.

\subsection{Regularization and Quantization}

\subsubsection{Space-time Lattice}
Quantum field theory usually is discussed within the perturbative
framework. Even there it is obvious that there are divergences in
elementary contributions. In a functional integration (Feynman path
integral) formulation one finds that the integration is not
well-defined. The functional integral has to be regularized in order to
be able to identify the singularities.

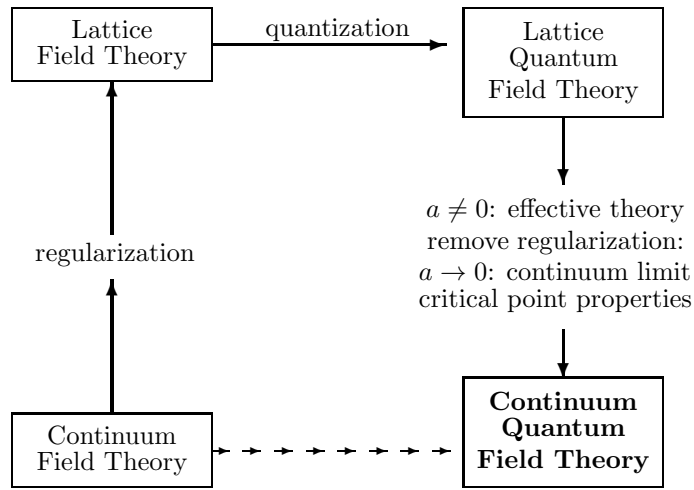
\unitlength=0.6mm
\linethickness{0.4pt}

\begin{center}
\begin{picture}(144.00,110.00)
\thicklines

\put(22,16){\vector(0,1){30}}
\put(22,56){\vector(0,1){34}}
\put(44,98){\vector(1 ,0 ){52}}
\put(122,82){\vector(0,-1){15}}
\put(122,35){\vector(0,-1){11}}

\thinlines
\put(0.00,0.00){\framebox(44.00,16.00)
{\shortstack{Continuum\\ Field Theory}} }

\put(0.00,90.00){\framebox(44.00,16.00)
{\shortstack{Lattice\\Field Theory}} }

\put(100.00,82.00){\framebox(44.00,24.00)
{\shortstack{Lattice\\ Quantum\\ Field Theory}} }

\put(100.00,0.00){\framebox(44.00,24.00)
{\shortstack{\bf Continuum\\ \bf Quantum\\ \bf Field Theory}} }
\put(5,50){\shortstack{regularization}}
\put(56,100){\shortstack{quantization}}
\put(90,40){\shortstack{
$a \neq 0$: effective theory\\
remove regularization:\\
$a \rightarrow 0$: continuum limit\\
critical point properties}}

\multiput(44,8)(7,0){8}{\vector(1,0){4}}
\end{picture}
\end{center}

Replacing the space-time continuum by a lattice provides a suitable
regularization as well as a computational tool \cite{Wi74}. We can
define the continuum QFT as the limit of the lattice QFT for vanishing
lattice spacings. Important symmetries of the Lagrangian (like e.g.
gauge invariance) may be sustained in the lattice formulation. On the
other hand, obvious symmetries of space-time like translational and
rotational invariance are replaced by their lattice subgroups and
eventual recovery in the continuum limit has to be established.

We will discuss euclidean QFT. The regularization replaces the
continuous, $O(4)$ symmetric space-time arena by a (in practical
calculations) finite lattice with symmetry $O(4;\Ga)$.
\begin{center}
\begin{tabular}{ll}
\hline
{}~&~\\
lattice&
$\Lambda$ e.g. $\Ga_N^D\subset \Ga^D$\\~&~\\
field&
$\Phi : \Lambda \mapsto \MX $  (e.g. $\Co$)
$\in \Gamma $\\~&~\\
configuration&
$\Gamma \in \MX^{\Lambda}$\\~&~\\
action&
$S_\Lambda : \Gamma \mapsto \RE$ \\~&~\\
partition function&
$Z_\Lambda = \int d\mu_\Lambda(\Phi)
\exp[S_\Lambda(g,\Phi)]$\\~&~\\
free energy&
$F_\Lambda = \ln Z_\Lambda$\\~&~\\
Gibbs factor&
$\rho_\Lambda (\Phi) = Z_\Lambda^{-1}
\exp[S_\Lambda(g,\Phi)]$\\~&~\\
expectation values&
$\langle A \rangle_\Lambda = \int d\mu_\Lambda(\Phi)
\rho_\Lambda (\Phi) A(\Phi)$\\~&~\\
\hline
\end{tabular}
\end{center}
Quantization thus amounts to determination of expectation values in a
4D statistical system, like e.g. the Ising model (cf. for instance
\cite{Ro91}).
With $g$ we denote the coupling constant (or a vector of constants)
parametrizing the action.

\subsubsection{Scaling}
The focus of interest in that process are two limits:
\begin{description}
\item{\bf Thermodynamic limit:} The infinite volume limit
$|\Lambda | \to \infty$.
\item{\bf Continuum limit:} Is obtained at a critical point
$g\to g_c$ in coupling constant space, where the
lattice spacing constant $a(g)\to 0$; the lattice spacing
acts as a regularization parameter.
\end{description}
The second limit needs some discussion. The physical quantities
like mass or length of the continuum theory should approach
constant values, however on the lattice they are expressed
in multiples of the lattice spacing $a$. These lattice quantities
are dimensionless and are  determined from the connected
correlation functions (propagators) of some operators
\bea
\langle A(0) A(x) \rangle_c &\propto& \exp (- x / \xi )
= \exp (- a x / a \xi )\nonumber\\
&=& \exp (- m_{\mbox{\footnotesize phys}} a x) \QA
\mbox{i.e.~} m_{\mbox{\footnotesize phys}} a = 1/\xi \; .
\eea
These correlation functions describe propagation in euclidean time and
the operators may be representing n-particle states. The exponential
decay should hold for sensible theories \cite{Se82}.

Only at a critical point the correlation length $\xi$
diverges\footnote{ At the critical point the correlation function decay
polynomially $\propto | x |^{d -2+\eta}$} and allows to let $a\to 0$
while keeping a fixed value for the physical mass parameter. One has to
fix the scale $a(g)$ by trading one  physical parameter. Once the scale
is set, all other parameters may then be expressed in terms of this
scale parameter $a(g)$.  However, all other physical parameters should
scale correspondingly, only then can we identify the correct continuum
theory.

In practical situations we distinguish between two qualities of scaling
behaviour.
\begin{description}
\item{\bf Scaling:} The dimensionless
ratios like ${m_1(g) a(g) \over m_2(g) a(g)} = {m_1(g) \over m_2(g) }$
scale towards the physical ratio $m_1 \over m_2$ as $a\to 0$
\item{\bf Asymptotic scaling:} The scaling behaviour
$a(g) \propto f(g)$ as
determined from the renormalization group $\beta$-function.
\end{description}

The relative importance of irrelevant operators determines the quality
of scaling. Whereas scaling may be observed quite early one has to be
quite close to the critical point in order to observe asymptotic
scaling.

\subsubsection{Continuum Limit}
Let us collect some points important in the context of the continuum
limit.
\begin{itemize}
\item The lattice action should obey Osterwalder-Seiler positivity
when approaching the continuum limit in
order to guarantee existence of a transfer matrix formalism and positive
norm Hilbert states \cite{OsSe78}.
\item The continuum limit is obtained at a critical point (2nd order
phase transition) of the lattice system: $g\to g^*$, where the
dimensionless correlation length $\xi \to \infty$.
\item All physical quantities are multiples of the lattice spacing and
should scale correspondingly, $a(g)\to 0$ for $g\to g^*$.
\item Lattice translation invariance $\mbox{mod}(a)$ should become
continuous translation invariance.
\item The space-time symmetry (cubic group for cubic lattices) should become
the continuous $O(4)$ (rotational) invariance. This may be observed in
off-shell observables (like the 1-particle propagator) as well as in
on-shell ones (like differential scattering cross-sections).
\item Fields, couplings and masses in the lattice action are bare
quantities and are renormalized in the quantization in continuum limit.
Renormalized quantities will have a subscript $r$ henceforth.
\item Because of renormalization, the continuum properties of a
quantized theory may be quite different from those of the bare theory.
In statistical physics this observation is linked to the notion of
universality: Theories with different micro-structure may show
universal critical (macroscopic) properties. Since the critical
properties are just the physical properties of our quantized continuum
theory this limits the class of liable continuum QFTs.
\end{itemize}

All these points should be observed in a sensible investigation of the
continuum limit. Most of these properties can be studied only in the
quantized theory in a non-perturbative environment.  In existing Monte
Carlo calculations several of these features have been checked and
some of these checks will be discussed in this series of lectures.

\subsection{Particle Physics and the $\Phi^4$ Model}

\subsubsection{Lattice Fields}
Typically we have to deal with the following types of lattice fields.
\begin{description}
\item{\bf Scalar bosons:} Occur in the action usually with local terms like
$\Phi_x$, $\Phi_x^2$, $\Phi_x^4$ and nearest neighbour terms like
$\Phi_x U_{x,\mu} \Phi_{x+\mu}$; the n.n. term is a lattice
discretization of the continuum derivative term. Monte Carlo integration
methods for these fields include the standard Metropolis or heat bath
updating as well as cluster- and multigrid methods. The computational
effort grows with $O(L^4)$ up to $O(L^6)$ depending on the algorithm ($L$
denotes the linear extension of the 4D lattice),
\item{\bf Gauge (vector) bosons: } Action terms include the Wilson
plaquette
interaction $Tr U_p$, where the plaquette variable is the path ordered
product of link fields $U_{x,\mu}$ around a plaquette. The gauge field
coupling $\beta \propto {1\over g^2}$ where $g$ is the usual
perturbative coupling constant. Monte Carlo updating algorithms include
Metropolis, heat bath and overrelaxation; some progress has been made in
designing multigrid algorithms. The computational effort grows with
$O(L^5)$ up to $O(L^6)$.
\item{\bf Fermions:} Occur in the action with terms like
$\overline{\Psi}_x
\Psi_x$ and $\overline{\Psi}_x Q_{xy}\Psi_y$ which includes Dirac
$\gamma$-matrices and nearest neighbour coupling $\kappa$. Updating
algorithms are an order of magnitude more demanding with regard to
computational resources; they include molecular dynamics updating,
Langevin and ``hybrid'' Monte Carlo methods as well as some recent
attempts towards multigrid methods. The central numerical problem is
the inversion of the lattice Dirac operator $Q$, thus the total effort grows
with
$O(L^9)$ up to $O(L^{16})$ depending on the algorithm.
\end{description}

For the formulation of the models on finite volumes one usually assumes
periodic boundary conditions, $\Phi(x+L\hat{\mu})=\Phi(x)$ (for fermions:
antiperiodic b.c.). The implicit assumption is that the thermodynamic
limit should be independent on the choice of boundary conditions.
Periodic b.c. seem to produce relatively small finite size effects.

\subsubsection{Scalar Model}
The standard model of electroweak and strong interactions describes
extremely successfully nowadays results in particles physics in an
energy domain from a few MeV up to more than 100 GeV. Its central
symmetry is $SU(2) \times U(1) \times SU(3)$. The $SU(3)$ is that of
QCD, represented by quarks, gluons and their interaction. The
non-perturbative lattice study of that model is a central topic in
lattice calculations \cite{De92}, but I will not try to cover it here.
The symmetry groups $SU(2) \times U(1)$ are those of quantum flavour
dynamics and are represented by the fundamental leptons, 4 gauge bosons
and the dublet $\Phi = (\Phi_1 + i \Phi_2, \Phi_3 + i \Phi_4) $ of
complex Higgs fields. In particular the action for the Higgs fields has
a symmetry $SU(2)_L\times SU(2)_R$ isomorphic to $O(4)$. Thus we may
represent the Higgs fields by a vector of 4 real scalar fields, too.

\FIG{StandardModel}{StandardModel.eps}{Standard model symmetry breaking
scenario.}{4.3cm}{4.5cm}{500}

In the standard scenario one assumes that there is a symmetry breaking
potential like the well-known Mexican hat potential $\lambda (\Phi^2
-1)^2$. In the Goldstone realization the vacuum chooses spontaneously
one direction (out of the infinite set of inequivalent
representations, cf. fig. \ref{StandardModel}),
thus one is left with one massive component (the
Higgs particle) and three massless components (the Goldstone modes),
which, in the so-called Higgs mechanism, become the longitudinal modes
of three of the (originally massless) vector bosons. The three vector
bosons acquire mass in that process. For sake of the later discussion
let us distinguish the two processes:
\begin{description}
\item{Symmetry breaking:} The $O(4)$ symmetry is spontaneously broken
and in the new vacuum we find one massive and three massless modes.
\item{Mass generation:} Three
massless gauge bosons buddy with the three Goldstone modes and become
massive gauge bosons, $Z$ and $W^\pm$.
\end{description}

Both mechanisms are non-perturbative as they involve infinitely many
degrees of freedom. The results mentioned are based on tree-level
perturbation theory and some additional assumption about the vacuum
structure. One of the basic questions to be answered in the
non-perturbative lattice simulations is, whether these results survive
quantization and whether they remain valid at arbitrary Higgs coupling
strength.

The essential non-perturbative ingredient is the $O(4)$ $\Phi^4$
model.  This is the main motivation for studying that theory
non-perturbatively in some detail.

\subsubsection{The $\Phi^4$ Model}
The continuum action
\be\label{HCaction}
S_{cont} = \int dx [\ot(\partial_\mu\Phi_x)^2
 + \ot m_{cont}^2\Phi_x^2
                +\frac{1}{4!} \lambda_{cont} (\Phi_x^2)^2
 + j_{cont} \Phi_x]
\ee
has three parameters, the bare mass, the 4-point coupling, and the external
field. Discretization is not unique and a particular simple form of a lattice
action is obtained by replacing the derivative term by a nearest neighbour
difference. With suitable redefinitions of the couplings we get
\be\label{SPhi4latt}
S_{latt} = \sum_\Lambda \bigl[-2\kappa \sum_\mu\Phi_x \Phi_{x+\mu} +
                + \lambda (\Phi_x^2-1)^2 + J \Phi_x\bigr] \; .
\ee
We have several important cases.
\begin{itemize}
\item $\Phi_x\in \Ga_2$: the Ising model,
\item $\Phi_x\in \RE$: the simple 1-component $\Phi^4$ model,
\item $\Phi_x\equiv\Phi^\alpha_x\in O(4)$: the 4-component $\Phi^4$ model, i.e.
the Higgs
model.
\end{itemize}
For notational simplicity we  sometimes use field variables
\be
\varphi \equiv \sqrt{2\kappa}\Phi\QA j\equiv J/\sqrt{2\kappa}\; .
\ee
In the form \eq{SPhi4latt} we identify immediately two limits of the
system. For $\lambda\to 0$ we approach a gaussian model. For $\lambda\to
\infty$ only configurations with $|\Phi_x|=1$ contribute and we
have a nonlinear sigma model (for the 1-component version: Ising model).
The coupling $\kappa, \lambda$ and $j$ in the lattice action are bare
quantities, given at the generic lattice scale $a$. Values measured in
the quantized system like e.g. the mass or the 4-point coupling are
renormalized quantities and will be denoted by a subscript: $m_r,
\lambda_r$.

At the critical point of the model, $\kappa=\kappa_c$,
critical exponents characterize the
singular behaviour of various quantities \cite{St71,Ma76a}.
Introducing the reduced coupling
\be
t\equiv {\kappa_c -\kappa \over \kappa_c}
\ee
we summarize in table \ref{TabCritInd} the critical behaviour and
definitions of the critical indices.  In sect. 2 we discuss the issue
of triviality of the model.  If it is indeed trivial (as all evidence
points to) the critical exponents are those of mean field theory and
there are non-leading logarithmic corrections. The exponents of these
corrections are given in table \ref{TabCritInd} and depend on $N$ of
the group $O(N)$ \cite{BrLeZi76,Ke93} ($N=1$ for the one-component
model).

\begin{table}
\caption{\label{TabCritInd}Summary of the expected critical behaviour
in the
$N$-component model in terms of the reduced critical coupling
$t\equiv (\kappa_c -\kappa)/\kappa_c$ and the external field $j$.}
\begin{center}
\begin{tabular}{lcll}
\\ \hline \\
$\mbox{\bf Order parameter}$\\
$\qquad \langle \Phi\rangle

            $&$ \sim
               $&$\displaystyle | t |^\beta
		|\ln |t||^\frac{3}{N+8} $&$\mbox{for}\; j=0$\\
$            $&$ \sim
               $&$\displaystyle j^{1/\delta}|\ln | j| |^\frac{1}{3}
                  $&$\mbox{for}\; t=0$\\
\\
$\mbox{\bf Susceptibility}$\\
$\displaystyle\qquad  \chi=\langle\Phi(p=0)^2\rangle_c
      =\frac{1}{|\Lambda|}\sum_x\langle \Phi_0\Phi_x\rangle_c
            $&$\sim
               $&$\displaystyle| t |^{-\gamma}|\ln |t||^\frac{N+2}{N+8}$\\
\\
$\mbox{\bf Internal (link) energy}$\\
$\displaystyle \qquad \langle E\rangle
= \sum_{x,\mu} \langle\Phi_x\Phi_{x+\mu}\rangle$\\
\\
$\mbox{\bf Specific heat}$\\
$\displaystyle\qquad  C_V= -\kappa^2 {\partial \langle E\rangle
\over \partial \kappa}
=\frac{\kappa^2}{|\Lambda|} (\langle E^2\rangle -\langle E\rangle ^2)
            $&$\sim
               $&$\displaystyle|t|^{-\alpha}|\ln |t||^\frac{4-N}{N+8}$\\
\\
$\mbox{\bf Correlation length}$\\
$\displaystyle \qquad \langle \Phi_0\Phi_x\rangle _c \propto
\exp (-m_r | x |)
\mbox{~for~}t\neq 0$\\
$\displaystyle\phantom{\langle \Phi_0\Phi_x\rangle _c} \propto
1/|x|^{d-2+\eta}
\mbox{~for~} t=0$\\
$\displaystyle\qquad  m_r = 1/\xi
            $&$ \sim
               $&$ \displaystyle |t|^\nu
                    |\ln |t||^{\frac{-N-2}{2N+16}}$ \\
\\
$\mbox{\bf Renormalized 4-pt coupling}$\\
$\displaystyle\qquad  \lambda_r\;\; = -{\langle\Phi(p=0)^4
\rangle_c\over\langle
\Phi(p=0)^2\rangle_c^2}$\\
$\displaystyle\qquad \mbox{~or~} =  \ot (m_r /\langle
\varphi_r\rangle )^2
            $&$\sim
               $&$\displaystyle| \ln |t||^{-1}$\\
$\displaystyle\qquad \mbox{~where~} \langle \varphi_r\rangle =
\langle \varphi\rangle
 /\sqrt{Z}$\\
{}~\\ \hline
\end{tabular}

\end{center}
\end{table}

The two definitions given for the renormalized charge agree in the
continuum limit (sometimes one uses the continuum definition action
with an extra factor 6:  $\lambda_r = 3  (m_r /\langle \varphi_r\rangle
)^2$).  Here $Z$ denotes the field renormalization constant (and will
be discussed in sect.3 in more detail) and not the partition function
(sorry, but thats how it is). Only two of the critical exponents are
independent (cf. e.g. \cite{Hu82}). The mean field values are for the
scaling exponent of the correlation length $\nu=\ot$ and for the
anomalous dimension $\eta=0$. From the scaling relations we find
$\alpha=2-d\nu=0$, $\delta={d+2-\eta \over d-2+\eta}=3$, $\gamma=\nu
(2-\eta)=1$ and $\beta=\gamma/(\delta-1)=\ot$.

\subsection{Monte Carlo Methods}

Quantization amounts to
determination of expectation values
\be
\langle A \rangle = \frac{1}{Z}\sum_{\Phi\in \Gamma} A(\Phi)
\exp[S(\Phi)]  \; .
\ee
In Monte Carlo integration one samples the field configurations with an
equilibrium probability distribution $P(\Phi)\simeq \exp[S(\Phi)]$. This
allows to determine expectation values according
\be
\langle A \rangle =\lim_{T\to\infty}\frac{1}{T}\int_0^T d\tau \;
A(\Phi(\tau))
=\lim_{N\to\infty}\frac{1}{N}\sum_{n=1}^N  A(\Phi_n).
\ee
The stochastic variable is the field $\Phi(\tau)$ evolving with the
stochastic (computer) time $\tau$. It is determined from a homogeneous
Markov process with transition probability $T$,
\be
P(\Phi_n=\varphi_n|\Phi_{n-1}=\varphi_{n-1})=T_1(\varphi_n|\varphi_{n-1})
\ee
i.e. $T_k$ depends only on the numbers of time steps $k$ but not on the
history. If $T_k(\varphi'|\varphi) > 0$ for all pairs $(\varphi',\varphi)$
and the process is aperiodic then it is ergodic.
A Monte Carlo integration algorithm usually
constructs such a Markov sequence of configurations, similar to a sequence
of snap shot of some realistic ferromagnet. Thus the reason for calling the
method a simulation. In realistic simulations ergodicity and the related
problem of relaxation are important questions. In particular if there are
topologically different sectors in configuration space some Monte Carlo
updating algorithms may have problems connecting them.

If implemented correctly the Monte Carlo integration leads to correct
results with a purely statistical error that decreases with the number of
iterations like $1/\sqrt{N}$. The results are for finite size lattices,
however. Control of both, the statistical error and the finite size effects,
belongs to the central issues of this field.

One starts with an (arbitrary) configuration. The Markov sequence of
configurations
\be
\varphi_0 \to \varphi_1\to \varphi_2\to \ldots
\ee
produces samples of the corresponding probability distributions, which,
depending
on certain sufficient conditions, approach the equilibrium distribution,
\be
P^{(0)} \to P^{(1)} \to P^{(2)} \to \ldots \to P\;\mbox{(equilibrium
distribution)},
\ee
where $P^{(n)}=T^nP^{(0)}$. In a calculation one therefore starts to measure
observables only after a sufficient number of ``equilibrating'' Monte Carlo
steps. Subsequent updates then produce configurations according to the
required distribution.

Meanwhile we have a collection of updating algorithms \cite{Wo92}. Some of
them proceed by changing the configurations locally, i.e. just one variable at
a time, others try to implement collective updates of nonlocal nature.
\begin{description}
\item{\bf Local updating algorithms}
\begin{itemize}
\item {\bf Metropolis}: The simplest and oldest \cite{MeRoTe53}
algorithm; we discuss it below.
\item {\bf Heat bath}: Like an iterated Metropolis updating, optimizing the
local acceptance rate.
\item {\bf Overrelaxation}: A sometimes very effective generalization of
the heat bath algorithm.
\item {\bf Microcanonical}: Reformulates the updating in terms of a
deterministic, discrete Hamiltonian evolution in a
space with doubled number of variables: $(\Phi, \Pi)$. A variant method
introduces a microcanonical demon.
\item {\bf Hybrid Monte Carlo}: Combines a number of microcanonical updates (a
``trajectory'') with a final Monte Carlo acceptance step.
\item {\bf Langevin}: Uses the stochastic differential equation for the
construction of configurations; equivalence to the microcanonical and the
hybrid method may be demonstrated (in certain limits).
\item {\bf Multicanonical}: Or multimagnetical or entropy sampling,
samples an ensemble with
modified energy density, particularly well suited for systems with
degenerate ground states (like at 1st order phase transitions or in
spin-glasses) \cite{BeNe91,BeNe92}.
\end{itemize}
\goodbreak
\item{\bf Non-local updating algorithms}
\begin{itemize}
\item {\bf Multigrid}: Employs scale changing transformations to identify
possible collective updating modes, essentially a simultaneous updating on
several length scales.
\item {\bf Cluster}:  Identifies the system-specific dynamical nonlocal
degrees of freedom; this method is very efficient for certain models and will
be discussed below in more detail.
\end{itemize}
\end{description}

\subsubsection{Metropolis Algorithm}
Considering a stochastic process $X(\tau)$, detailed balance is a sufficient
condition that the probability distribution approaches the equilibrium
distribution $\exp{[S(x)]}$,
\be
P_1(x| y)e^{S(y)} = P_1(y| x)e^{S(x)} \Rightarrow
\lim_{\tau \to \infty} P(X(\tau)=x) \simeq e^{S(x)} .
\ee
Summation over $x$ shows that the limiting distribution is an eigenvector
of $P_1$. The Metropolis algorithm \cite{MeRoTe53}, starting at some value
$x$ (i.e. some initial configuration), consists of the steps:
\begin{enumerate}
\item Choose some $y$ according to some {\em a priori\/} transition
probability $P_T(y| x)$ .
\item Accept the new value $y$ with the probability
\bea
P_A(y | x) &=& \min \left( 1, \frac{P_T(x| y) \exp[S(y)]}{P_T(y|
x) \exp[S(x)]} \right)\\
&=& \min \bigl( 1, \exp[S(y)-S(x)] \bigr)\mbox{ for } P_T(x|
y)=P_T(y| x)\; .
\eea
It is straightforward to see, that the resulting $P_1= P_T P_A$ fulfills
detailed balance. In particular for symmetric $P_T$ the information
necessary to decide on acceptance or rejection comes only from the change
of the action $\Delta S$ with regard to the change of the variable. If this
change is local, i.e. just at some field variable $\Phi_x$, the change of
action may be determined from the field values in the local neighbourhood.
\item Repeat these steps from the begin.
\end{enumerate}
One MC iteration is one sweep through the lattice, that is one attempt to
update for each field variable on the lattice. Typical calculations perform
$O(10^6)$ MC iterations for given lattice size and couplings constant values.

The exponential autocorrelation time controls the statistical dependence of
subsequent configurations and may be defined
\be
|P^{(n+1)} - P| \leq \exp{(-1/\tau_{exp})} |P^{(n)} - P|
\ee
or from the autocorrelation of some observable A,
\be
\Gamma_A(n)=\langle A(\varphi_0)A(\varphi_n)\rangle - \langle A\rangle^2 \simeq
\exp{(-n/\tau_{exp,A})}\QA\tau_{exp,A}\equiv \sup_A\tau_{exp,A}\;,
\ee
or from the integrated autocorrelation time
\be
\tau_{int,A} = \ot + \sum_{n>0} {\Gamma_A(n) \over \Gamma_A(0)}\; .
\ee
The latter quantity controls directly the statistical error in estimates
of $\langle A\rangle$; a run of $n$ measurements contains
only $n/2\tau_{int,A}$ effectively independent measurements.
This relaxation behaviour affects particularly systems with large
correlation lengths $\xi$ (near critical points), since the autocorrelation
length
grows like
\be
\tau_{exp} \simeq \min(L,\xi)^{z}\QA \tau_{int,A} \simeq \min(L,\xi)^{z_A}
\ee
where $z$ and $z_A$ denote the dynamical critical exponents. They depend
on the updating algorithm and have the value $2$ for most local algorithms
like the Metropolis algorithm (for overrelaxation algorithms this value may
be smaller: $\sim 1$ \cite{Wo92}). The computational effort for the simulation
of a
system of linear extension $L$ near a critical point\footnote{For systems
with several coexisting phases the autocorrelation length grows even
exponentially with $\simeq \exp (c L^{d-1})$;
multicanonical algorithms address this
tunneling problem.}, where $\xi\simeq L$, grows like $L^{d+z_A}$. This
phenomenon is called {\em critical slowing down\/}. For more
detailed discussions cf. \cite{So89,So91,Wo92}.

In the MC simulation the statistical error of the measured quantities
shrinks with the inverse square root of the number of statistically
independent configurations. Improvement of the performance therefore can
be achieved only by improving the updating algorithm by making them
computationally simple (and quick), by reducing the autocorrelation time
and by reducing the dynamical critical
exponent $z$. This is the motivation for the introduction of multigrid- and
cluster updating algorithms.

\subsubsection{Cluster Algorithm}
Fortuin and Kastelyn \cite{FoKa69,FoKa72} pointed out that the Ising model has
an
alternative formulation as a bond percolation model. This leads to clusters
of connected bonds \cite{CoKl80}, which are the nonlocal dynamical objects
of the Ising model.

Swendsen and Wang \cite{SwWa87} suggested a very efficient updating
algorithm based on alternating between the spin and the cluster
representation. For a given spin configuration one first identifies
Ising clusters built of neighboured sites with parallel spins. Some of
the bonds are then deleted with probability $\exp (-2\kappa)$, the
resulting bond clusters are the dynamical FK-CK-SW-Clusters
\cite{FoKa69,CoKl80,SwWa87}. With a probability 0.5 one now can flip
simultaneously all spins in a cluster, repeating this for all clusters.
This results in a new spin configuration and the updating step is
completed (cf. fig. \ref{Clusterupdating}).

\FIG{Clusterupdating}{SW_algorithm.eps} {The Swendsen-Wang cluster
updating algorithm:  $\{s\}\to\{b\}\to\{b'\}\to\{s'\}$.}
{7cm}{6.5cm}{600}

It turns out that this algorithm gives $z\simeq 0.3$ for the 2D and
$z\simeq 1$ for the 4D Ising model. Wolff proposed a modification,
where one randomly chooses a site and identifies the cluster connected
to it like in the SW-algorithm; this  cluster is then flipped
unconditionally. This results in $z\simeq 0$ for the 4D system!

The cluster algorithm originally was designed for systems like the
Ising or the Potts model, where the spin variables take values in a
discrete group.  Wolff \cite{Wo89} observed, that one may extend its
scope of application to continuous groups, that have embedded Ising
variables (the group manifold may be split into disjoint parts, cf.
\cite{So91}), like e.g. the group $O(N)$.  This allowed to utilize the
updating method for $N$-component $\Phi^4$ theories. The cluster method
has provided a major breakthrough, the efficiency of the algorithm has
brought forward results for scalar models with highest computational
precision.

The cluster representation allows the construction of observables with
reduced variance, so-called {\em improved estimators\/}
\cite{Sw83}. The propagator between spins at two
sites, e.g., may be obtained by just counting how often both spins belong to
the some cluster:
\be
\langle \phi_x \phi_y \rangle = \langle \sum_{\mbox{\footnotesize all clusters
c}}
\chi_c(x)\chi_c(y) \rangle\quad \mbox{where}\quad
\chi_c(x) = \left\{
\begin{array}{lcl}
1 &\mbox{ if }& x \in c \\0& \mbox{ if }& x \notin c\end{array} \right.
\; .
\ee
For a given set of configurations this allows determination of the
propagators for larger distances with much better statistical
quality than achieved from the direct measurement.

\FIG{FractalCluster}{redclu.eps}
{A 3D cross section through a large cluster of a 4D Ising system
near the critical point}
{10cm}{9cm}{600}

The clusters are the dynamical objects of the critical system, a realization
of Fisher's droplets. One therefore would expect relations between the
geometrical properties and some critical indices. Due to scale invariance at
a critical point clusters should look alike on different length scales. In
quantum mechanics this is expressed by the fractal structure of the
quantum paths, which are of Brownian nature. We might expect such a
fractal structure also in the field configurations of quantum field theory
(cf. fig. \ref{FractalCluster}).

\FIG{fractaldimension}{fractal_dimension.eps}
{In this log-log plot \protect{\cite{JaLa91}}
of observed combinations $(s,R)$
of clusters at the critical point the intermediate scaling region becomes
obviously larger for larger lattices. Its slope defines the fractal dimension.}
{10cm}{7cm}{800}

The relation between such a fractal structure and the odd critical index may
be argued as follows \cite{St79}. Under a scale transformation we have
scaling of the reduced critical temperature $t$ (cf. table \ref{TabCritInd})
in the form
\be
\xi \rightarrow b \xi\QA t \rightarrow b^{-1/\nu} t
\ee
where $\nu$ denotes the scaling  index of the correlation length. In a volume
$L^d$ the total magnetization scales according to table \ref{TabCritInd}
\bea
M(L) &\propto &L^d t^\beta\quad \rightarrow\\
M(bL) &\propto &(bL)^d b^{-\beta/\nu}t^\beta = b^{d -\beta/\nu} M(L)\; .
\eea
We therefore expect finite size scaling of $ M(L)\simeq L^{d -\beta/\nu}$,
an anomalous dimension. This scaling behaviour
should be followed by the clusters, as
they are relevant for the critical behaviour. They should therefore exhibit a
fractal dimension $d_f=d -\beta/\nu=d/(1+1/\delta )$. In 4D we
expect mean field critical exponents $\delta=\frac{1}{3}$ and
thus $d_f=3$.

In a computer study on a finite lattice the clusters are of limited extension.
Small clusters are essentially one-dimensional lattice animals, whereas
large clusters will feel the boundaries and fill the space and then behave
like d-dimensional objects. Scaling behaviour we can expect only at
intermediate size clusters. However, the larger the lattice, the larger the
domain of intermediate size. In a study of the 4D $\Phi^4$ theory with
$O(4)$ symmetry \cite{JaLa91} the relation between mass $s$ (number of
sites belonging to the cluster) and size $R$ (the square root of the mean
square radius between points of the cluster) was studied.

In fig. \ref{fractaldimension} results of simulations for 4D lattices of
various size demonstrate the existence of a scaling region. Its slope is the
fractal exponent in the relation
\be
s \propto R^{d_f}
\ee
with a value $d_f\approx 3.05$ indeed compatible with the expectations.
An earlier study for the 3D Ising model \cite{StWa90} had produced a
value $d_f=2.51$ also compatible with the expected relation to
the corresponding critical exponents in $d=3$.

\section{Triviality of $\Phi^4$}
\setcounter{equation}{0}

\subsection{Rigorous Results}

The claim ``This theory is trivial'' should be formulated more
explicitly:  Defining the continuum theory as the limit $a\to 0$ of the
lattice theory, then all $n$-point functions may be expressed through
products of 2-point functions (a theory of generalized free fields),
which even may be those of a free particle (a theory of free fields).

Let me summarize here the main results.
For more details one should refer to a review like \cite{GlJa87,Ca88a}.

\begin{description}
\item{$D<4$:} The continuum limit defines an interacting field theory. The
renormalized 4-point coupling $\lambda_r$ is bounded
($0\leq\lambda_r\leq const$) and perturbation theory applies
(cf. \cite{GlJa75,GlJa76,BrFrSo83} and further references in
\cite{Ca88a}).
\item{$D>4$:}
The continuum limit defines a trivial theory with mean field behaviour; all
connected $2n$-point functions (Wightman-distributions) vanish. Most
likely this is a theory of free fields (cf. \cite{Ai81,Fr82,BrFrSp82}).
\item{$D=4$:}
No rigorous proof of triviality exists, however there is a collection of
results which leave little room for another result.
\begin{itemize}
\item Triviality has been conjectured by Wilson \cite{Wi71,WiKo74}
in the framework of perturbative renormalization group  and, under this
assumption, many scaling results have been derived (cf. \cite{BrLeZi76}).
\item The renormalized 4-point coupling $\lambda_r$ is bounded
($0\leq\lambda_r\leq const$); for the single component theory the
critical exponents are bounded from below by gaussian (mean field) values
 \cite{GlJa87}.
\item Considering the equivalence to random walks \cite{Sy66,Sy69} led to
exact inequalities like $\lambda_r\leq 3 p_{is}$ (where $p_{is}$
denotes the intersection probability of 4D random walks)
 \cite{Ai81,Ai82,AiGr83,Fr82,ArCaFr83}.
\item Rigorous control of block spin transformations at the critical point
(the massless theory) proves triviality for {\em small enough\/}
$\lambda_r$ \cite{GaKu85b}; close to the
critical point there are logarithmic corrections \cite{Ha87a,HaTa87}.
\end{itemize}

If the continuum limit is trivial and therefore defined by a fixed point of
gaussian nature then its nature does not change in the complete $SU(2)$
gauge Higgs model \cite{HaHa86} -- although another
additional FP at some different value cannot be excluded.
{\em Assuming triviality\/} one can
find regions, where the high temperature series (expansion in $1/\kappa$)
still gives reliable results and the (renormalization group) scaling
behaviour is already valid. This overlap allows to continue $n$-point
functions from the symmetric (hot) phase through the phase transition
 \cite{LuWe89}. Triviality implies a bound on the renormalized coupling and
on the Higgs mass, even if the model is just an effective
theory \cite{CaMaPa79,DaNe83}. We will discuss these issues below in
more detail.
\end{description}

\FIG{PhaseDiagram}{phi4_phase_diagram.eps}
{Phase diagram of the 1- or (in parentheses) 4-component $\Phi^4$
model}
{7cm}{8cm}{500}

The phase diagram of the $\Phi^4$ model in the $(\kappa, \lambda)$ plane is
sketched in fig.\ref{PhaseDiagram}. For the 1-component model one has a
massive Higgs boson in both phases. At the phase transition the
dimensionless mass vanishes due to $a\to 0$. The 4-component model
has four degenerate massive bosons in the symmetric phase. In the broken
phase one finds one massive boson and three massless Goldstone modes
(although $a>0$). The figure also gives regions of applicability of the
standard perturbative and non-perturbative methods. Monte Carlo
calculations cover the whole domain of interest but they are limited to
finite volume. On the other hand, the analytic methods are limited to certain
domains and have to assume dominance of the gaussian fixed point (GFP). A
thorough analytic understanding, however, helps to analyze the Monte Carlo
results in a reliable way, as I try to demonstrate below.

Before we turn to Monte Carlo methods and results let us briefly review the
results of renormalization group. Perturbation theory gives the
$\beta$-function that describes the rescaling properties of the coupling
with respect to a change of the scale $\mu$. For the 1-component model the
one-loop result is
\be
\left( \partial \lambda \over \partial \ln \mu\right) = \beta(\lambda)
\equiv\frac{3}{16 \pi^2} \lambda^2 + O(\lambda^3)\; .
\ee
Solution of the differential equation gives
\be
\lambda (\mu) = {\lambda(\Lambda) \over
1 - \lambda(\Lambda) \frac{3}{16 \pi^2} \ln
(\mu/\Lambda)}\quad\mbox{or}\quad
\lambda (\Lambda) = {\lambda(\mu) \over
1 + \lambda(\mu) \frac{3}{16 \pi^2} \ln (\mu/\Lambda)}\; .
\ee
The cutoff $\Lambda$ corresponds to the inverse lattice spacing $1/a$,
thus $\lambda(\Lambda)=\lambda_{bare}$. For the physical scale $\mu$ one
could choose some low-energy mass (like the Higgs mass), thus
$\lambda(\mu)=\lambda_r$.  Taking these results at face value, for
fixed $\lambda(\mu)$ the cutoff $\Lambda$ cannot be pushed to arbitrary
high values due to a zero in the denominator (Landau pole). Or, to rephrase
this problem as a question: How large do we have to make the microscopical
coupling $\lambda_{bare}$ in order to have a non-zero macroscopical
coupling $\lambda_r$?

\FIG{CouplingBound}{phi4_RG_bound.eps}
{The renormalization of the quartic coupling as expected from
renormalization group arguments.}
{7cm}{5.5cm}{500}

In fig.\ref{CouplingBound} this situation is summarized. Trajectories of
fixed bare coupling are bounded from above by the $\lambda_{bare}=\infty$
values of $\lambda_r$. Holding $\lambda_r$ fixed but attempting to
raise the cutoff (approach the critical point, i.e. the continuum limit) is
not possible - one gets stuck at the upper bound at some value of $\Lambda$
(or, equivalently, at some non-zero value of $a(\kappa) > 0$. The bound
forces the renormalized coupling to approach zero in the continuum limit.
Only at finite cutoff non-vanishing values of $\lambda_r$ are
permitted. Thus even a trivial theory may have an effective
interaction below the cutoff.

This  discussion is based on the one-loop perturbative result and the
renormalization group equations. It was one of the challenges of the Monte
Carlo approach to confirm (or not confirm) this scenario. Although the
asymptotic shape of the upper bound is given, the multiplicative factor is of
non-perturbative nature. Also, the shape at small values of $\Lambda$ is
regularization dependent. The corresponding amplitudes are universal
up to regularization dependent corrections $O(\mu^2/\Lambda^2)$.

In a sensible regularization the cutoff should be larger than the masses of
the physical states, thus $\Lambda/\mu > 1$. If the situation is like
discussed above, this provides an upper bound
on $\lambda_r$. Perturbation theory in the electro-weak gauge coupling $g$
relates the renormalized coupling with the ratio between the Higgs mass
$m_H$ and the mass of the heavy vector boson $W$,
\be\label{EqHiggsBound}
\lambda_r = \ot R^2 \QA R = {m_H\over F}
= {g \over 2} {m_H\over m_W} \QA F\equiv \langle\varphi\rangle /\sqrt{Z}\;.
\ee
(As mentioned, sometimes the continuum notation with an extra
factor 6 is used: $\lambda_r = 3 R^2$, cf. table \ref{TabCritInd}.)
Given the experimental values
($F\simeq 246$ GeV, or $m_W$ = 80.6(4) GeV, $g^2\simeq 0.4$)
the non-perturbative upper bound on
the renormalized 4-point coupling provides an upper bound on the Higgs
mass \cite{CaMaPa79,DaNe83}, unrelated to the usual unitarity-bound.

What evidence can we expect from Monte Carlo simulations? There are three
groups of results
\begin{enumerate}
\item[(a)] Phase structure and critical exponents, derived from bulk
quantities like the internal energy, susceptibility or specific heat; detailed
investigation of the finite size properties of the partition function allows
quantitative statements on finite size scaling, the thermodynamic limit and
logarithmic corrections.
\item[(b)] Monte Carlo renormalization group (MCRG) methods allow to determine
the flow of coupling and the fixed point structure, including the critical
exponents.
\item[(c)] Propagator measurements lead to the mass spectrum and values
of the renormalized couplings. These results are of particular interest
concerning the triviality bound on the Higgs mass and its regularization
dependence. The energy spectrum also allows the determination  of phase shifts.
\end{enumerate}

In the subsequent discussion of this section I first present more
qualitative results from real space renormalization group,  then a high
precision determination of the partition function and the logarithmic
corrections. In sect. 3 the non-perturbative results on the bound are
discussed.

\subsection{Real Space Renormalization Group}

In perturbative renormalization group the number of couplings
considered is very small, although the number of couplings generated in
the transformation may be infinite. The Monte Carlo renormalization
group method, also called real space renormalization group (RSRG)
method since it works in position space, allows to deal with a larger
coupling space, although one has to truncate as well.

The {\em Scaling Hypothesis\/} suggests, that the singularities of the
thermodynamic quantities are due to the divergence of the correlation
length \cite{Fi72,Ba83,Ca88}. At the critical point the system should
be scale invariant and under scale changing transformations
\be
Z(\Lambda) \stackrel{T(\Lambda, \Lambda')}{
\begin{picture}(35,6) \put(0,2){\vector(1,0){35}}\end{picture}}
 Z'(\Lambda')
\ee
the physical properties (i.e. the properties at long
distance in terms of lattice spacing) and thus the partition function should be
invariant. Here the transformation $T$ is element of a semigroup. On a
lattice it is realized by the Wilson-Kadanoff
{\em block-spin transformation\/} \cite{WiKo74}, where the local
degrees of freedom are ``integrated out''. We have
\begin{center}
\begin{tabular}{llcl}
scale transformation & $\Lambda$ & $\rightarrow$ & $\Lambda'$ \\
linear factor $b$ & $|\Lambda| /|\Lambda'|$ &=& $b^D$\\
invariance & $Z(\Lambda)$  &=&$Z'(\Lambda')$\\
field variables & $\Phi\in\Gamma$&$\rightarrow$& $\Phi'\in\Gamma'$\\
correlation length & $n_\xi a$ & $\rightarrow$ &$ n_\xi' a' =
( n_\xi /b) b a$\\
\end{tabular}
\end{center}
For lattice systems $T$ transforms an ensemble of configurations on
$\Lambda$ into an ensemble of configurations on $\Lambda'$.
The correlation length changes by the scale factor $b$.

Given
that the original ensemble was distributed according to a Gibbs measure
for an action $S(\kappa,\Phi)$ let us for the moment assume that the
blocked ensemble also follows such a measure, but for a different
action $S(\kappa',\Phi')$ (cf. the discussion further down).
Let us formulate the BST (block spin transformation) $\Phi \to \Phi'$
by introducing a
normalized transformation probability $P$  with the properties
\begin{enumerate}
\item[(a)] normalization: $\int d\mu_{\Lambda'}(\Phi')  P(\Phi', \Phi)=1$
\item[(b)] positivity: $P(\Phi', \Phi)\geq 0$; with (a) this implies
that for all $\Phi$ there is a $\Phi'$ such that $ P(\Phi', \Phi)\neq
0$, i.e. there is a non-vanishing probability of mapping into
each sector of $\Gamma'$.
\item[(c)] symmetry: all symmetries of the action $S$ should be respected.
\end{enumerate}
We then may describe the transformation by
\bea
Z(\Lambda) &=& \int d\mu_{\Lambda}(\Phi)\exp[S(\kappa,\Phi)] =
 \int d\mu_{\Lambda}(\Phi)  d\mu_{\Lambda'}(\Phi')  P(\Phi', \Phi)
\exp[S(\kappa,\Phi)] \nonumber \\
&\equiv& \int  d\mu_{\Lambda'}(\Phi')  \exp[S(\kappa',\Phi')] =
Z'(\Lambda')
\eea
with
\be
\exp[S(\kappa',\Phi')] = \int d\mu_{\Lambda}(\Phi)  P(\Phi', \Phi)
\exp[S(\kappa,\Phi)]\; .
\ee
The BST is largely arbitrary \cite{BuLe82} and may be chosen optimal
from the point of view of simulational simplicity. One has considered
both, deterministic and probabilistic forms,
\be
\begin{array}{rcll}
\displaystyle P(\Phi',\Phi) &=& \displaystyle\prod_{x'}
\delta(\Phi_{x'}', f_{x'}(\Phi))
&\mbox{(deterministic)},\\
 &=&\displaystyle
 \prod_{x'} \exp(\Phi_{x'}' \sum_\alpha\rho_\alpha f^{\alpha}_{x'}(\Phi))
&\mbox{(probabilistic)}.
\end{array}
\ee
A deterministic transformation
with a linear relation between the spins and the blockspins makes
the new spin at $x'$ an average over the old spins within some block
$B_{x'}$,
\be
f_{x'}(\Phi) \propto \sum_{x\in B_{x'}} \Phi_x \; ,
\ee
whereas a nonlinear form could be chosen to keep certain expectation
values (in the example below: $\langle\Phi^2\rangle $) fixed during the
transformation,
\be
f_{x'}(\Phi) = \mbox{sign}[ \sum_{x\in B_{x'}} \Phi_x ]\; \frac{1}{16}
 \sum_{x\in B_{x'}} \Phi_x^2 \; .
\ee
For Ising spin systems this choice gives the majority rule. Probabilistic
transformations have adjustable parameters $\rho_\alpha$ that may be
used to optimize certain features.

Under repeated transformations the (dimensionless) correlation length
shrinks and the equivalent action changes.
\bea
n_\xi \rightarrow n_\xi' \rightarrow n_\xi'' \rightarrow&\ldots&
n_\xi^* = \left\{
\begin{array}{ll}
0 &\mbox{trivial FP}\\
\infty& \mbox{critical FP}\end{array}\right.\; , \\
\kappa \rightarrow \kappa' \rightarrow \kappa'' \rightarrow&\ldots&
\kappa^* \quad\mbox{(fixed point)}\; .
\eea
The couplings renormalize according
\be \label{couplingrenorm}
\kappa'=\kappa'(\kappa)
\ee
which serves as the definition of the BST in action space, and,
assuming simple FP (fixed point) properties, $\kappa^*=\kappa'(\kappa^*)$.
If we start at a critical point, the correlation length is infinite all
the way: One stays on the critical hypersurface. If the correlation
length was finite it will approach zero in the iteration of the BST,
corresponding to the approach to a trivial (hot or cold) FP.

At the FP we may linearize \eq{couplingrenorm} to obtain
\be
\kappa_\alpha'(\kappa)=\kappa_\alpha^* +
T_{\alpha\beta}^*(\kappa-\kappa^*)_\beta +
o(||\kappa-\kappa^*||) \quad\mbox{with}\quad
T_{\alpha\beta} =
\left({\partial \kappa_\alpha' \over\partial
\kappa_\beta}\right)_{\kappa^*}
\ee
or, for the n-th step
\be
T_{\alpha\beta}^{(n+1,n)} =
{\partial \kappa_\alpha^{(n+1)} \over\partial\kappa_\beta^{(n)}}
\; \longrightarrow \; T_{\alpha\beta}^* \; .
\ee
The eigenvalues $\{\lambda\}$ of the FP values of the linearized
BST decide on the number of relevant ($\lambda>1$), marginal
($\lambda=1$), or irrelevant ($\lambda<1$) operators (couplings)
at the critical FP. (There may also be redundant operators. They occur
e.g. if the set of couplings is overcomplete \cite{MuSh85,GuPa87}).
Fig. \ref{RGflowplot} shows an example for the RG flow. Starting at the
FP with a small perturbation, the flow will lead away into the direction
of relevant operators. This defines the renormalized trajectory, along
which there are no contributions of irrelevant operators, i.e. no
corrections to scaling.

\FIG{RGflowplot}{RG_flow_diagram.eps}{Schematic display of the
renormalization flow in coupling space; starting from the vertical
axis at $\kappa_2=0$}{7cm}{5cm}{600}

Usually almost all couplings (corresponding operators in the action)
are irrelevant. The attraction domain of a FP is high-dimensional
and only a few (two or three) coupling are relevant. This behaviour is
the reason for universality classes: The same critical surface is domain
of attraction for many different initial actions. Although the position
of the critical points may be different, the critical behaviour is not,
as it is always given by the properties of $T^*$.

The linearized BST $T^*$ factorizes into even and odd subspaces,
depending on the symmetry properties of the corresponding terms in the
action under reflection $\Phi\leftrightarrow -\Phi$. The relevant
eigenvalues are related to the corresponding critical exponent.
This may be seen from the scaling behaviour of e.g. the correlation
length. Assuming that its critical behaviour is given by some
critical index $\nu$ we find
\bea
\xi(\kappa)&=&b\xi'(\kappa'(\kappa))\rightarrow \nonumber\\
(\kappa - \kappa^*)^{-\nu} &= &
b \left( \kappa'(\kappa) - \kappa^* \right)^{-\nu}
=b \lambda^{-\nu} (\kappa - \kappa^*)^{-\nu}
\eea
which is consistent only for
\be
\nu = {\ln b \over \ln \lambda}\; .
\ee
If there are marginal eigenvalues there is a line of FPs in the
critical surface.

The position of the FP on the critical surface may depend on details
of the BST \cite{Sw84a}. In particular it may be moved into
directions of redundant operators \cite{FiRa86}.  In practice
the amount of redundancy is often not clear.  Improvement may be
desirable for both, the FP action (cf. \cite{GuPa87}) and the
renormalized trajectory \cite{Sw84a,GaLa87,LaSa88}.
Recent attempts to optimize the BST such that the FP action becomes
particularly simple and as local as possible seem to
be promising for asymptotically free theories \cite{HaNi93}.

Before I continue, let me mention possible difficulties and caveats.
\begin{itemize}
\item The transformation may introduce spurious singularities as
discussed by Griffiths and Pearce \cite{GrPe78,GrPe79}.
\item Some BSTs at first order transitions may produce non-Gibbsian
measures (\cite{EnFeSo91,EnFeSo93}, cf. however \cite{MaOl93}).
\item The FP may describe a nonlocal action (no exponential
damping of nonlocal terms) or it may be at infinity in the
parameter space of the action.
\item Depending on the lattice geometry and dimensions only a limited
range of blocking factors is available.
\item Like in perturbative renormalization group the number of couplings
generated in the transformation may be infinite. In practical
calculations one therefore has to rely on truncations.
\end{itemize}

For the following discussion we write the action as a sum of
operators,
\be
S = \sum_\alpha \kappa_\alpha S_\alpha \;.
\ee
Expectation values and correlators are related to derivatives of the free
energy
\be
{\partial F \over \partial \kappa_\alpha} = \langle S_\alpha\rangle
\QA
{\partial^2 F \over \partial \kappa_\alpha\partial \kappa_\beta} =
{\partial \langle S_\alpha\rangle \over \partial \kappa_\beta} = \langle
S_\alpha S_\beta\rangle-\langle S_\alpha\rangle\langle
S_\beta\rangle \; .
\ee
BSTs provide different possibilities to obtain information on the
critical properties.
\begin{description}
\item{\bf Critical exponents:} The cross-correlations between observables
measured on the original and the blocked configurations allow to get
estimators for the truncated $T$-matrix \cite{Ma76,Sw79}.
\bea \label{LinTfromBST}
{\partial\langle S_\alpha^{(n+1)}\rangle\over \partial \kappa_\beta^{(n)}}&=&
\sum_\gamma {\partial\langle S_\alpha^{(n+1)}\rangle
\over \partial \kappa_\gamma^{(n+1)}}
{\partial\kappa_\gamma^{(n+1)}\over \partial \kappa_\beta^{(n)}} =
\sum_\gamma {\partial\langle S_\alpha^{(n+1)}\rangle
\over \partial \kappa_\gamma^{(n+1)}}
T_{\gamma\beta}^{(n+1,n)} \; ,\\
{\partial\langle S_\alpha^{(n)}\rangle \over \partial \kappa_\beta^{(m)}}
&=& \langle S_\alpha^{(n)}S_\beta^{(m)}\rangle -
\langle S_\alpha^{(n)}\rangle \langle S_\beta^{(m)}\rangle \; .
\eea
Here  $T$ is obtained from \eq{LinTfromBST} by inversion. Due to
the truncation there will be
inherent errors (see however \cite{ShGuMu85}). This approach has been
very successful for 2D and 3D spin models where it has provided the maybe
best estimates of critical exponents \cite{BuLe82a}.
\item{\bf Flow in the space of observables:} Repeated BST leads one to the
renormalized trajectory. Matching of observables allows to relate
couplings of different simulations.
\item{\bf Flow in the space of couplings:} This would allow to identify
FPs and universality class in a very intuitive and direct way.
\end{description}

Unlike in perturbation theory, in non-perturbative RSRG calculations it
is no straightforward task to measure renormalized couplings. It
is simple to construct the blocked configurations (in position space),
but one has no direct means to know about the corresponding action.
There is the problem to learn the values of (many, in the thermodynamic
case infinitely many) couplings of the blocked action, which is not
known directly.  A possibility is by operator matching: Matching the
observables determined by Monte Carlo simulations on lattices of the
size of the blocked lattice \cite{ShTo80,HiSc83,HaHaHe84,BoHaHa86}.
However, for this one has to combine statistically independently
determined expectation values which needs high statistics. Another,
very efficient method has been suggested by Swendsen \cite{Sw84} and
amounts to determine conditional expectation values in the background
of given configurations.

MCRG calculation for the one component $\Phi^4$ model have been
performed on lattices up to size $16^4$ and the results were compatible
with the expected scenario: one relevant even and one relevant odd
direction in operator space. The result for the even critical exponent
was good ($\nu = 0.53(4)$ from e.g. \cite{BlSw80,La86c}, or from
\cite{CaPe84a}) but the statistical accuracy not sufficient to identify
the expected next-to-leading marginal eigenvalue. No convincing signal
for logarithmic corrections could be determined in those calculations.

However, from the flow structure of the renormalized couplings one
could try to find out, whether different values of the bare quartic
coupling are in one universality class. This was done in \cite{La86c}
in a space 24 different couplings corresponding to 6 quadratic (local
and distance 1, 2, 3, and 4 spin-spin products) and 18 quartic (local
and neighbouring 2-, 3- and 4-spin products) operators. The simulation
was done on a $16^4$ lattice with most of the initial couplings set to
zero. Only the nearest-neighbour coupling $\kappa$ and the bare 4-point
coupling $\lambda$ were non-zero. For given fixed values of $\lambda$
(0.2, 0.4 and 2) the model was simulated for the critical value of
$\kappa$ and BSTs were performed down to lattices of size $8^4$ and
$4^4$. The renormalized couplings in the 24D space were determined with
Swendsen's method \cite{Sw84}.

\FIG{LambdaFlow}{LambdaFlow.eps}{Flow structure in coupling space;
starting on the critical surface in the $(\kappa_1,\kappa_2)$ plane
(corresponding to the bare quartic coupling and n.n. coupling)
the RG flow
moves towards a common FP, independent of the bare quartic coupling: it
is irrelevant.}{7cm}{5cm}{500}

The observed flow may be summarized as in fig. \ref{LambdaFlow}.  From
different starting values of the bare quartic coupling on the critical
surface the flow is towards a common FP. In the study different BSTs
were considered. As the FP position may depend on the particular BST
one could expect different flow structure, which indeed was observed
\cite{CaPe84a,La85a,La86c}. The deterministic nonlinear type of BST
discussed above produced a flow towards the FP characteristic for the
Ising system. The overall result demonstrated, that the bare quartic
coupling is not relevant for the FP and thus for the continuum theory
-- in agreement with the notion of triviality.

\subsubsection{Critical Indices from Bulk Quantities}
Bulk quantities like the internal energy per link $\langle E\rangle $
or the magnetization $\langle\Phi\rangle$ are the most direct way to
the determination of scaling behaviour. However, they are affected by
finite size effects and therefore the critical exponents may be masked
behind the smooth crossover between the phases typical for finite
systems. Finite size scaling \cite{Bi92,Ba83} relates finite size
properties with the critical exponents in the thermodynamic limit.
However, usually one needs very precise determination of the quantities
involved. The histogram method, an old idea \cite{FaMaPa82} reshaped
into the multi-histogram approach by Ferrenberg and Swendsen
\cite{FeSw88,FeSw89,FeSw89a}, allows to combine measurements at
different couplings in an efficient way.

We rewrite the partition function in terms of the density of states
(for simplicity we assume that the set of energy values is discrete,
otherwise we have to rely on some binning of the allowed range).
\bea
Z(\kappa) &=& \int d\mu_\Lambda(\Phi) \exp [\kappa S(\Phi)] = \int
d\mu_\Lambda(\Phi)\sum_E \delta(S-E) \exp [\kappa S(\Phi)] \nonumber\\
&=&\sum_E \rho(E) \exp (\kappa E) = \sum_E p(E|\kappa) \; .
\eea
The
probability for a configuration having the total action $E$ for
coupling $\kappa$ is $p(E|\kappa)$ and is estimated by the
corresponding histogram values $h(\kappa,E)$ obtained in the Monte
Carlo simulation of the system. Simulations at different values of the
coupling provide different estimators for the spectral density
$\rho(E)$:
\bea
\kappa_0&\rightarrow&\rho_0 (E)=h(\kappa_0,E)\exp (-\kappa_0 E)\; ,
\nonumber \\
\kappa_1&\rightarrow&\rho_1 (E)=h(\kappa_1,E)\exp (-\kappa_1 E)\QA \mbox{ etc.}
\eea
However, the statistical accuracy for these different estimators will
be best for those values of $E$ close to the corresponding expectation
value (i.e. near the peak of the corresponding histogram).  A suitable
combination of histograms \cite{FeSw88,FeSw89,FeSw89a}, however, gives a
reliable estimator over a broader range of energies (and therefore
couplings). From this estimator for the spectral density one may
construct energy histograms at arbitrary values of the coupling, as in
fig. \ref{SpectralDensity}.

\FIG{SpectralDensity}{SpectralDensity.eps}{The logarithm of the
probability
distribution (histogram) as reconstructed from the multihistogram:
$\ln[\rho(E)e^{\kappa_0 E}]$ vs. $E$, for $\kappa_0=0.15055$ and
lattice size $16^4$.}{10cm}{7cm}{800}

This technique may be generalized to other observables and terms in the action.
Taking into
account energy and magnetization we have the partition function
\be \label{SpecRepZ}
Z(\kappa, h) = \sum_{E,M} \rho(E,M)\exp (\kappa E + h M)
\ee
allowing the precise determination of expectation values of arbitrary functions
of $E$ and
$M$,
\be
\langle f(E,M)\rangle = \frac{1}{Z} \sum_{E,M} \rho(E,M) f(E,M) \exp (\kappa E
+ h M) \; .
\ee
An example is the specific heat $C_V$; in fig. \ref{SpecHeat} both, the
values determined at specific couplings and the continuous curve from
the multi-histogram analysis is shown. Each curve contains all the
information of the corresponding individual couplings (for that lattice
size).

\FIG{SpecHeat}{SpecHeat.eps}
{The specific heat $C_V$ for the one-component $\Phi^4$
model determined for  lattice sizes $8^4$, $12^4$, $16^4$, $20^4$,
and $24^4$ (from \protect{\cite{KeLa91}}).}{8cm}{5.5cm}{480}

The representation \eq{SpecRepZ} also allows for analytic continuation
to complex values of $\kappa$ and $h$ since only the exponential has to
be continued. One therefore can determine the zeroes of $Z$ in the
complex $\kappa$ and $h$ planes. In the actual calculation this can be
done reliably only close to the values of the couplings, where the
original histograms have been determined, otherwise the statistical
errors of the estimated spectral density blow up and lead to large
uncertainties. Experience shows, that extrapolation is feasible in an
elliptical domain with the real positions of measurements between the
foci of the principal axis.

The study of zeroes of the partition function has been started by Lee
and Yang \cite{YaLe52,YaLe52a}; zeroes in the complex external field $h$
are called Lee-Yang zeroes. They have to lie at purely imaginary values
of $h$ (on the unit circle in the fugacity $z=e^{-2h}$
plane) \cite{YaLe52,YaLe52a}. Zeroes in the even coupling (the
temperature) have been studied in particular by Fisher \cite{Fi68} and are
thus usually called Fisher zeroes. On a finite lattice the partition
function is a finite-degree polynomial in $e^{-4\kappa}$ and
$z$.

\FIG{PartitionZeroes}{nul.eps}{{\bf Left:}
Finite size scaling
behaviour of the imaginary part of the closest Fisher-zero $\kappa_1$
for lattice size $L=8$, $12$, $16$, $20$ and $24$ at $h=0$. (a) The
leading exponent is related to the slope of the straight line and a fit
gives -2.088(6) slightly differing from the mean field expectation
$-1/\nu=-2$. This difference is due to the presence of logarithmic
corrections as demonstrated in (b) where the leading behaviour has been
cancelled by considering $L^2 \mbox{Im} \kappa_1$ .  {\bf Right:}
Finite size scaling behaviour of the imaginary part of the closest
Lee-Yang-zero $h_1$ for lattice size $L=8$, $12$, $16$, $20$ and $24$
and $\kappa=0.149703$. (a) The leading exponent is related to the slope
of the straight line and a fit gives -3.083(4) slightly differing from
the mean field expectation $-1/\delta=-3$. This difference is due to
the presence of logarithmic corrections as demonstrated in (b) where
the leading behaviour has been cancelled by considering $L^3 \mbox{Im}
h_1$ (figures from \protect{\cite{KeLa93}}).} {13cm}{9cm}{400}

Expressing, e.g. $Z$ through a product of zeroes we find, that the
specific heat -- the second derivative of its logarithm -- is peaked
for those real values of $\kappa$ that are closest to a nearby complex
zero:
\be
Z\propto \prod_i (\kappa-\kappa_i) \quad \rightarrow \quad
C_V(\kappa) \propto \sum_i\frac{1}{(\kappa-\kappa_i)^2}\; .
\ee
In the thermodynamic limit the zeroes accumulate and pinch the real
$\kappa$-axis (and, for critical coupling,  the real point $h=0$): They
build the critical singularity. The $\Lambda$-dependence of the
positions of the zeroes closest to the real coupling axis can be
related to the thermodynamic critical exponents (finite size scaling)
and the logarithmic corrections.

Assuming that the gaussian FP controls the leading critical behaviour,
perturbative renormalization group methods may be applied to predict
the critical properties \cite{BrLeZi76} and the finite size scaling
behaviour of the partition function zeroes \cite{Br82}.  For the
one-component $\Phi^4$ model this was done recently
 \cite{KeLa91,KeLa93,KeLa93a} (for the $O(N)$ model cf. \cite{Ke93}) with
the following results for the Fisher zeroes (using the reduced coupling
$t$) and the Lee-Yang zeroes:
\bea \label{FiScaling}
| t_j| &\sim & L^{-2} (\ln L)^{-\frac{1}{6}}\; ,\\
 \label{LYScaling}
| h_j| &\sim & L^{-3} (\ln L)^{-\frac{1}{4}}\; .
\eea
Here the leading exponents are the negative inverse of the mean field
values for $\nu$ and $\delta$, the multiplicative logarithmic
corrections depend on the number of components of $\Phi$ and have
other values for the $N$-component model. This result also gives the
finite size scaling behaviour for the peak values of the specific heat
and the susceptibility
\bea
c_L(t=h=0)&\sim & (\ln L)^{\frac{1}{3}}\; , \\
\chi_L(t=h=0) &\sim & L^2 (\ln L)^\ot\; .
\eea

In a high statistics simulation the multi-histogram techniques were
used to study bulk quantities of the $\Phi^4$ model and the positions
of the closest partition function zeroes \cite{KeLa91,KeLa93,KeLa93a}. The
numerical result together with the Ferrenberg-Swendsen technique
allowed sufficient precision to identify both, the leading critical
exponent and the form of the logarithmic correction.  The results for
the Lee-Yang and for the Fisher-zeroes (cf. fig.
\ref{PartitionZeroes}) confirmed the expectation. The leading even and
odd exponents agree with the mean field result. The
logarithmic corrections are in good agreement with the expressions
\eq{FiScaling} and \eq{LYScaling}
derived on ground of perturbative renormalization group.

\section{Consequences of Triviality}
\setcounter{equation}{0}

Let us now concentrate on results derivable under the impact of
triviality like the bound on the renormalized coupling and
the Higgs mass.  We therefore discuss in this section in particular the
$O(4)$ model because of its close relationship to the unified theory of
electro-weak interaction. The notation refers to the lattice action
\eq{SPhi4latt} given above.

\subsection{Results from Non-Monte Carlo Techniques}

Using a local potential approximation and Wilson recursion relation
technique for renormalization group calculations Hasenfratz and Nager
\cite{HaNa88} obtained analytic and non-perturbative results on the
bound for the one-component model. The effect of the inherent
approximations is hard to judge in that approach.  A real
break\-through, however, were the results of L\"uscher and Weisz
\cite{LuWe87,LuWe88,LuWe89} and for this reason I want to briefly
summarize them.

The authors study the $N(\geq1)$-component model on a hypercubic lattice.
Starting from the symmetric phase they proceed as follows (cf. fig.
\ref{PhaseDiagram}):

{\noindent \bf Symmetric phase: }
The high temperature (small $\kappa$) expansion for the renormalized mass
$m_r$, coupling $\lambda_r$ and the critical value $\kappa_c$ is determined
up to $O(\kappa^{14})$. The values seem to be reliable up to relatively small
values of  $m_r\simeq 0.5$, already within the scaling region.

{\noindent \bf Scaling (critical) region: }
Here one uses the results of perturbation theory in the renormalized
quantities:  $m_r$ and $\lambda_r(\kappa, \lambda)$ for fixed bare
$\lambda$, using 3-loop $\beta$-functions. One has to assume that the
gaussian FP defines the universality class. The a priori unknown
integration constants like $C_1$ in
\be \label{ScalingMassRen}
\kappa<\kappa_c : a m_r(\lambda) = C_1\,(\lambda)(\beta_1
\lambda_r)^{-\beta_2/\beta_1^2}\exp\left(-\frac{1}{\beta_1 \lambda_r}\right)
\, \{1 +O(\lambda_r)\}
\ee
contain the non-perturbative information and are taken from the
$\kappa$-expansion values in the overlap region. Corrections to scaling are
$O(a^2(\ln a)^c)$. The scaling form is used to continued through the phase
transition to the phase with broken symmetry in a region $m_r\leq 0.5$.

{\noindent \bf Broken symmetry phase: }
The Integration constants below and above the critical point are related
through the massless theory at the critical point,
\be
\kappa>\kappa_c : C_1'(\lambda) = e^\frac{1}{6} C_1(\lambda) \; .
\ee
They are renormalization prescription dependent.

\FIG{LuWeRGFlow}{nul.eps}{Quantitative plot of renormalization group
trajectories for the $O(4)$ model (from \protect{\cite{LuWe89}}). The
abscissa variable $\bar{\lambda}$ corresponds to values of the bare
coupling $0\leq\lambda\leq\infty$. The range of $\kappa$-values roughly
covers the scaling region $| m_r| \leq 0.5$. The curves correspond to
constant renormalized 4-point coupling -- they approach the critical
line, but they never touch it: In the continuum limit one is forced to
vanishing values of $\lambda_r$.}{5.5cm}{7cm}{500}

This program exploits standard analytic techniques in a very efficient
way and worked nicely \cite{LuWe87,LuWe88,LuWe89}. In fig.
\ref{LuWeRGFlow} trajectories of constant renormalized 4-point coupling
are shown: They never cross the critical line, therefore only a
non-interacting continuum limit is feasible.

We summarize the restrictions or assumptions for this approach:
\begin{itemize}
\item the $\kappa$-series is truncated;
\item triviality (gaussian FP) is assumed;
\item validity of perturbation theory in the renormalized couplings
(renormalization group at the 3-loop level) is assumed.
\end{itemize}
Direct Monte Carlo simulations can check these assumptions and confirm the
validity of the approximations. They also can determine further results
concerning, for instance, the regularization dependence. A consistent picture
gives weight to both, the analytical and the numerical approach.

\subsection{Monte Carlo Simulations}

First results go back to 1984 \cite{Wh84,Ts85} but stayed widely
unnoticed. Only with the very high statistics studies of various groups
 \cite{HaJaLa87,MoWe87,KuLiSh88,KuLiSh88b,BhBi88,HeNeVr92} since 1987 the
results
became
convincing enough. These studies worked with lattice sizes up to $16^4$. Most
results have been obtained in the broken phase, both for $j=0$ and $j>0$.
Typically several 100.000 sweeps and measurements per point have been
performed, using first the Metropolis, later the cluster algorithm. In summary
thousands of hours supercomputer CPU-time (like e.g. Cray-YMP or equivalent)
have been spent for these analyses. Some studies were concerned with the
one-component model but, again, most concentrated on the $O(4)$ model.

For the determination of the renormalized coupling in the phase of broken
symmetry we need (cf. \eq{EqHiggsBound} and table \ref{TabCritInd})
\begin{itemize}
\item The vacuum expectation value $\Sigma=\langle \varphi\rangle$ ; the
complication is,
that for $j=0$ in finite volumes there is no spontaneous symmetry breaking
and one has to consider suitable definitions.
\item The field renormalization constant $Z$ or, equivalently, the {\em pion
decay constant\/} $F\equiv \Sigma/\sqrt{Z}$. This may be determined e.g. from
the propagators of the massless Goldstone modes.
\item The renormalized mass $m_r$ of the massive (Higgs) mode, determined
from the measured Higgs propagator.
\end{itemize}
In addition to these quantities  necessary for the derivation of $\lambda_r$ we
may also want to identify clear signals for the existence of Goldstone modes,
like
e.g. propagators at $j=0$ and $j>0$, just to be sure that the symmetry
breaking mechanism is the conjectured one.

The problem to define, what one means by spontaneous symmetry
on finite lattices, turned out to be hard in practice. A straightforward
definition
originally employed was
\be \label{Sigmalocal}
\Sigma = \langle |
\overline{\varphi}^\alpha|\rangle\quad\mbox{where}\quad
\overline{\varphi}^\alpha = \frac{1}{|\Lambda|} \sum_x\varphi_x^\alpha ,
\ee
and $m_r$, $F$ and $Z$ were determined from the propagators of the field
components parallel and perpendicular to the value
$\overline{\varphi}^\alpha$ of
the corresponding configuration, which meandered around in the $O(N-1)$
subspace. That seemed to work nicely.  In fig. \ref{MCsigmamass} we show
results for the Higgs mass determined that way on lattices of various
size. Further
results showed that the field renormalization constant stays close
to the value 0.97 from the PT up to  $m_r\approx 1$.

\FIG{MCsigmamass}{nul.eps}{The dimensionless renormalized Higgs mass
$a(\kappa)m_r$ determined on lattices of various size and extrapolated to
infinite
lattice size (from \protect{\cite{Ne89b}}).}{5.5cm}{5.5cm}{1000}

A better understanding of the infinite volume limit of the {\em ad hoc\/}
ansatz \eq{Sigmalocal} was important to achieve systematically improved
quality of the results for $\Sigma$. Now, for $j\neq 0$ the symmetry is
broken explicitly, all quantities are well defined and can be measured
without reference to spontaneous symmetry breaking. For small values of
the external field the would-be Goldstone states are still very light
and therefore they feel the finiteness of the volume best. We have a
situation of two different kinds of long correlation lengths (fig.
\ref{TwoCorrLengths}). At the 2nd order PT $a(\kappa) m_r \rightarrow
0$, because $a(\kappa)\rightarrow 0$ for $\kappa\to\kappa_c$, since the
theory is critical. Along the 1st order PT $a(\kappa) m_G(j)
\rightarrow 0$ because the Goldstone boson mass $m_G(j) \rightarrow 0$
for $j\to 0$. In the discussion below we will denote masses of the massive
mode $m_\sigma$ and of the Goldstone mode $m_\pi$ in analogy to
the nonlinear $\sigma$-model.

\FIG{TwoCorrLengths}{TwoCorrLengths.eps}{Phase structure in the $(\kappa,
j)$-plane.}{6.5cm}{3.3cm}{600}

Let us recall again the important parameters of the model. The field
$\Phi^\alpha$ has four real components. In the symmetric
(high temperature) phase
the particle spectrum has 4 degenerate
states. In the low temperature phase $\kappa>\kappa_c$,
at infinite volume $O(4)$ symmetry is spontaneously broken.
In this phase
\be
\langle\varphi_x^0\rangle =\Sigma\QA
\langle \varphi_x^i\rangle =0 \quad (i=1,2,3)\; .
\label{SIGMA}
\ee
Here the spectrum of the theory contains three
massless Goldstone bosons corresponding to the excitations
in the O(4) directions $\alpha = i = 1,2,3$.
Therefore, the correlation functions do not fall off exponentially
at large distances, but with an inverse power of the distance.
Specifically, the Goldstone boson two-point function satisfies
\be
\lim_{| x-y |\to \infty}
4\pi^2 | x-y |^2\langle \varphi^i_x\varphi^j_y\rangle =Z\delta^{ij} .
\label{Z}
\ee

For $O(4)$ symmetry the model contains six currents, all conserved at
$j=0$, whose charges generate the group.  The corresponding Ward
identities strongly constrain the behaviour of the correlation
functions at large distances.  In fact, the asymptotic behaviour of all
the Green functions associated with the currents and with the fields
$\varphi^\alpha_x$ is determined by the two low energy constants
$\Sigma$ and $F$ \cite{GaLe84}.  In analogy with QCD, $F$ is often
referred to as the pion decay constant.  It specifies the matrix
elements of the axial currents $A_\mu^j(x)$ (in Minkowski space)
between the ground state and single Goldstone boson states,
\be
    \langle 0| A_\mu^j(0) |\pi^k(p) \rangle = i\delta^{jk} p_\mu F\; .
\label{F}
\ee
The low energy constants $Z$, $\Sigma$ and $F$ are related through
\be
       F=\Sigma/\sqrt{Z}\; .
\label{ZSIGMAF}
\ee

For $j>0$ the Goldstone boson mass $m_\pi$ has a non-vanishing
value which in the lowest order of chiral perturbation theory is
\be
         m_\pi^2 = j\frac{\Sigma}{F^2}\; .
\label{MGB}
\ee

Whenever (almost) massless states dominate the system the low energy
(long distance) structure is of a universal character determined by the
symmetries of the dynamics in terms of a few low energy constants and
by the geometry of the finite system.  One therefore may study an
effective field theory with only Goldstone modes and the symmetry
structure of the underlying model $O(N)\to O(N-1)$
\cite{Le87,GaLe88,HaLe90,Ne88}.  The effective Lagrangian has the same
symmetry as the original system but is simple enough to permit a {\em
systematic\/} perturbative expansion. It has the form
\bea
L_{\mbox{\footnotesize eff}}&=& \ot F^2 (\partial_\mu \vec{S}\partial_\mu
\vec{S})
- \Sigma \vec{H}\vec{S} \nonumber\\
 &&+ \mbox{further terms of higher order in $\vec{H}$
and $\vec{S}$}\; ,\\
Z(\vec{H})&=&\int d\mu (\vec{S}) \exp (\int dx
L_{\mbox{\footnotesize eff}})\; .
\eea
In higher orders of this perturbative Lagrangian some further low
energy constants $\Lambda_{\Sigma}$, $\Lambda_F$ and $\Lambda_M$ are
required \cite{HaLe90}. They are the scale parameters determining the
logarithmic dependence of $\Sigma$, $F$ and $m_\pi$ on $j$.  In systems
of finite size $L$ this expansion amounts to an expansion in powers of
$L^{-2}$.  It is reliable because the interaction between the Goldstone
modes is weak at low momenta.

The first application of this idea \cite{GaLe87,Le87} has been made in
the context of the study of finite size effects in QCD with light
quarks, whose low energy properties are determined by the chiral
SU(2)$\otimes$SU(2)$\simeq$O(4) symmetry \cite{GaLe84}.  Therefore one
refers to this approach  as ''chiral perturbation theory''. A
comprehensive description can be found in \cite{GaLe88,Le88,HaLe90}.

The effective theory and its expansion relates values of quantities
determined at finite volumes and  $j>0$ with those at $j=0$ at infinite
volume,
\be
\Sigma(j,V), F(j,V), Z(j,V) \stackrel{\mbox{\small Chiral Pert. Th.}}{
\begin{picture}(45,6) \put(0,2){\vector(1,0){45}}\end{picture}}
\Sigma, F, Z .
\ee

\FIG{UVexpansion}{nul.eps}{The typical domains of application for two
type of expansions indicated by lines of constant $m_\sigma / m_\pi$ in
the $(\kappa,j)$ plane.  The dashed curve corresponds to $m_\pi L=1$ on a
$10^4$ lattice.  The regions below (above) the dashed curve are the
$\cal U$ ($\cal V$) domains for $L =$ 10 (from
\protect{\cite{HaJaJe91}}).}{5.5cm}{5.5cm}{1000}

There are two types of expansions, valid in different domains with a
non-vanishing overlap (cf. fig. \ref{UVexpansion}).
\begin{description}
\item[($\cal U$) $ m_\pi\lsim 1/L$ ~:~]
This domain is characterized by very small symmetry breaking, where
the symmetry is restored.
The finite size effects are large here.
The expansion is in powers of $L^{-2}$ keeping the total magnetic
energy
\be \label{U0}
u_0 = \Sigma j L^4\; ,
\ee
fixed. Thus $j$ is treated as a quantity of the order $O(L^{-4})$.
In $\cal U$ the correlation length $\xi_\pi$ grows
as $L^2$ for $L\rightarrow\infty$.

\item[($\cal V$) $m_\pi\gsim  1/L$ ~:~]
In this region $j$ is still small, although larger than in $\cal U$.
Finite size effects are smaller than in $\cal U$.
A typical range observed is $m_\sigma/m_\pi \gsim 5-10$.
Here the symmetry is not restored.
The expansion in powers of $L^{-2}$ in that domain keeps fixed
\be
           v_0 = \Sigma j L^2/F^2 = m_\pi^2 L^2 \; .
\label{V0}
\ee
\end{description}

For large $u_0$ the results of the expansion $\cal U$ should
smoothly go over into those of the expansion $\cal V$.
There is an overlap of the regions of validity
of both expansions around $m_\pi L\simeq 1$ \cite{HaJaJe91}.
In fig. \ref{UVexpansion} the dashed curve indicates where the
$\cal U$ and $\cal V$ regions in the $\kappa$-$j$ plane meet for $L = 10$.

We do not want to go into details of the various expansions. Just in
order to catch the spirit of the approach let us give an example for one
result of the $\cal U$-expansion.
\be
\langle \varphi^0_x\rangle_{V,j}= \frac{u}{jL^4}\frac{ I_2(u) }{ I_1(u)}
         +2 \rho_2\frac{\Sigma^2}{F^4}j+O(L^{-6})  ,
\label{PHILJ}
\ee
where the second term is $O(L^{-4})$ and
\be
\frac{u}{jL^4} =
\Sigma [1 + \frac{3\beta_1}{2L^2F^2} +O(L^{-4})+O(L^{-6})] .
\ee
The second term here is $O(L^{-2})$. The constant
$\beta_1=0.14046$ depends on the lattice geometry, as does the constant
in $\rho=0.01523 + 0.0095\,O(\ln L\Lambda_\Sigma)$.
The terms $O(L^{-4})$ are small and involve higher orders in
the chiral expansion.

\FIG{SigmajL}{nul.eps}{Example for the type of data obtained for
$\langle \varphi^0_x\rangle_{V,j}$ (from \protect{\cite{HaJaJe91}}).
The solid
curves correspond to a fit according \protect{\eq{PHILJ}} up to
$O(L^{-2})$, the resulting value of the only parameter $\Sigma$
is indicated by the full dot at $j=0$.}{9.5cm}{6cm}{1000}

Eq. \eq{PHILJ} gives the functional form of the $j$ and $L$
dependence in terms of the thermodynamic quantities $\Sigma$ and $F$.
Fig. \ref{SigmajL} gives an example of the type of data and fit quality
obtained. The leading term in \eq{PHILJ} provides the value of
$\Sigma$. The non-leading term in principle can give $F$, but is
statistically not sufficiently reliable. For this parameter one better
uses the t-dependence of the Goldstone propagator.

\FIG{SigmaPlot}{nul.eps}{The values of $\Sigma$ obtained from the
fit of $j>0$, finite volume data to the chiral perturbation
expression; the curve shows the fit to the expected scaling
behaviour (from \protect{\cite{HaJaJe91}}).}{5.5cm}{5.5cm}{1000}

The values $\Sigma(\kappa)$ obtained in this way prove to be very
precise, as demonstrated in fig. \ref{SigmaPlot}. One should keep in mind,
that these {\em are the thermodynamic values\/}! The fit to the scaling
behaviour gives \cite{HaJaJe91}
\be
\Sigma^2 = 0.672(6) (-t) |\ln (-t)|^\ot
\mbox{~and~}\kappa_c=0.3036(10) .
\ee

In table \ref{SigmaTable} the values of $\Sigma$ obtained by the
discussed chiral perturbation expansion method \cite{HaJaJe91} is
compared with the original ``brute force'' ansatz \eq{Sigmalocal} and
one finds surprising agreement - although the new method definitely is
of higher quality.

\begin{table}
\caption{Results on various quantities of the $O(4)$ $\Phi^4$ model
as obtained with the action \protect{\eq{SPhi4latt}} in various
studies.\label{SigmaTable}}
\begin{center}
\begin{tabular}{llllllll}
\hline
$\kappa$
& $\Sigma$
& $\langle | \overline{\varphi}^\alpha|\rangle$
& $F$
& $Z$
& $m_\sigma$
& $m_\sigma/F$
& Ref. \\
\hline
0.3130(7)& 0.196(1)    &    &0.199(5)    &0.976(6)   &0.50      &2.517(65)
& \cite{LuWe89}\\
0.3101(5)& 0.166(6)   &    &0.168(5)    &0.974(7)   &0.40      &2.380(69)
& \cite{LuWe89}\\
0.3077(3)& 0.130(3)   &    &0.132(3)    &0.973(5)   &0.30      &2.280(58)
& \cite{LuWe89}\\
0.3058(2)& 0.093(2)   &    &0.094(2)    &0.972(5)   &0.20      &2.129(46)
& \cite{LuWe89}\\
0.3046(1)& 0.051(1)   &    &0.052(1)    &0.971(5)   &0.10      &1.941(34)
& \cite{LuWe89}\\
{}~\\
0.355 & 0.4036(9)  & 0.402  & 0.4109(13) & 0.965(2)  & 1.09(10)  & 2.65(25)
& \cite{HaJaJe91} \\
0.330 & 0.3027(11) & 0.301  & 0.3075(16) & 0.969(3)  &  0.92(7)   & 2.99(24)
& \cite{HaJaJe91} \\
0.325 & 0.2769(12) & 0.276  & 0.2807(19) & 0.973(5)  &  0.81(2)   & 2.89(9)
& \cite{HaJaJe91} \\
0.310 & 0.1643(63) & 0.163  & 0.1668(73) & 0.97(1)   &   0.39(2)  & 2.34(22)
& \cite{HaJaJe91} \\
0.3075 & 0.132(13) & 0.128  & 0.135(14)  & 0.96(1)   &   0.29(1)  & 2.15(30)
& \cite{HaJaJe91} \\
{}~\\
0.3100   & 0.15549(10)&  & 0.1584(7)  &0.964(7)   &0.4073(7)  & 2.572(15)
& \cite{GoKaNe93} \\
0.3080   & 0.12827(6) &  & 0.1301(3)  &0.972(3)   &0.3223(12) & 2.477(14)
& \cite{GoKaNe93} \\
0.3060   & 0.09045(10)&  & 0.0915(2)  &0.977(3)   &0.2072(7)  & 2.264(14)
& \cite{GoKaNe93} \\
\hline
\end{tabular}
\end{center}
\end{table}

Let us summarize the different ways to determine the low energy
parameters.

{\bf Determination of $\Sigma$:} (a) from the $j>0$ data with help of the
discussed chiral perturbation expansion in $j$ and $L^{-2}$;
(b) at j=0 from the distribution of the ``local'' mean magnetization
$\langle |\overline{\varphi}^\alpha|\rangle$ defined in
\eq{Sigmalocal} -- for a detailed discussion of this constraint
effective potential approach cf.
 \cite{Go90,GoLe91,GoKaNe92,GoKaNe93}.

{\bf Determination of $Z$:}  from the Goldstone propagator (a) at $j>0$
from chiral perturbation expansion formulae \cite{HaJaJe91} and (b) at
$j=0$ using the field components transversal to the direction defined
by the  magnetization $\langle |\overline{\varphi}^\alpha |\rangle$.

{\bf Determination of $m_\sigma$ :} (a) from the component of the field
parallel to the direction of $\overline{\varphi}^\alpha$ (for each
configuration); (b) from the susceptibility
\be \chi = |\Lambda|
\left(\langle |\overline{\varphi}^\alpha|^2\rangle -\langle
|\overline{\varphi}^\alpha|\rangle^2          \right)\; ,
\ee
which is
in leading order proportional to $1/m_\sigma^2$. The latter method
appears to work very efficiently \cite{GoKaNe93}.

\FIG{UpperBoundNeuhaus}{nul.eps}{Comparison of results
for the renormalized coupling for different values of the
cutoff $\Lambda = 1/(am_\sigma)$ (from \protect{\cite{Ne89b}} in 1990). }
{10cm}{6cm}{1000}

\FIG{UpperBoundAachen}{UpperBoundAachen.eps}{Results for the
renormalized coupling for different values of the cutoff $\Lambda =
1/(am_\sigma)$. The 4 points with larger error bars are from
\protect{\cite{LuWe89}} and correspond to twice the error quoted,
following their discussion of uncertainties. The upper group of points
corresponds to a different action as will be discussed below (from
\protect{\cite{GoKaNe93}} in 1992).} {9cm}{6.5cm}{400}

In table \ref{SigmaTable} we compare the results derived over the
years from various groups. One observes increasing quality of the
numbers.  Comparing the Monte Carlo results for the renormalized
coupling at values $m_\sigma<0.5$ with those of
\cite{LuWe87,LuWe88,LuWe89} one finds good agreement (within the
statistical errors). This was also confirmed by further Monte Carlo
results of other groups \cite{KuLiSh88,KuLiSh88b,BhBi88}. The resulting
estimates for the upper bound on the renormalized coupling are shown in
fig. \ref{UpperBoundNeuhaus} and fig. \ref{UpperBoundAachen}. One
clearly notices the progress in the statistical accuracy.

If we assume that the cutoff (given in units of $m_\sigma$ in the
figure) should not be below $\sim 2$ then we read off an upper bound of
$R\lsim 2.6(3)$. With \eq{EqHiggsBound} this translates into $m_H\lsim
$670(80) GeV. This bound comes from the field theoretic properties of
the $\Phi^4$ model as a direct consequence of triviality and should not
be mixed up with the so-called unitarity bound on the Higgs mass.  The
bound expresses the fact, that for this part of the standard model we
cannot perform the continuum limit. Or, put differently, for a given
Higgs mass there is a limited domain of validity of the regularized
$\Phi^4$ theory. If the Higgs mass comes out to be much smaller, e.g.
of the order of 200 GeV, the bound allows a cutoff of the order of the
Planck length. Up to the allowed cutoff we still can consider the
theory a model with an effective interaction. However, we conclude that
the theory cannot be a truly fundamental one (we cannot remove the
cutoff completely) and we therefore may expect constituent structure.

All these results have been obtained away from the critical point,
therefore one does have contributions from non-leading operators.
These corrections to scaling are regularization dependent and
$O(a^2)$.  The cutoff then represents new, possibly more fundamental
physics which would replace the standard-model at higher energies.
However, its influence is effective on all scales and manifests itself
in the cutoff dependence of all observables.  In describing the
experiments through a lattice model, the higher in energy we go the
stronger we expect such cutoff effects to be. More and more terms would
have to be included in the effective action in order to explain
experiments with sufficient accuracy.  Within the scaling region one
expects to find these dependencies to be only weak, but this knowledge
is mainly based on rough estimates and there are only few explicit
studies \cite{BhBi88,BhBiHe90,He90,GoKaNe92,GoKaNe93}.

Value and shape of the bound depend on the adopted regularization
scheme, i.e. on:
\begin{itemize}
\item The lattice geometry; an investigation of the scalar field theory
on various types of lattices is required \cite{BhBi88,Ne87,Ne90,He90}.
Methods for comparing the results obtained by means of different
regularization schemes are being developed
\cite{Ha89,LuWe89,La89}; they require calculations of high
precision in a region of the coupling parameter space where the universal
asymptotic scaling associated with the (''trivial'') gaussian fixed
point sets in.
\item Additional terms in the action; an example is the
introduction of a straight distance 2 coupling in  order
to remove the $O(p^4)$ dependence in the lattice perturbative
propagator (Symanzik improvement).
\item Other terms of $O(1/\Lambda^2)$ in the action; Heller
et al. \cite{HeKlNe92,HeKlNe93a,HeNeVr92} have studied
these effects in the $O(N)$-model by systematically adding operators of
increasing order in $1/\Lambda$, both, in continuum and on the lattice,
in particular on an $F_4$-lattice \cite{Ne87}.
\item Other regularization schemes: Recent work
demonstrates the regularization dependence in the continuum $O(4)$
model with a higher derivative regulator \cite{JaKuLi93,JaKuLi93a}.
\end{itemize}

Comparing results from different actions obtained at finite cutoff is
not straightforward, though. Different actions have different scaling
violation properties. Choosing couplings such that the observed
correlations lengths agree is not the only and not necessarily the best
definition.  Consider e.g. an effective action for block variables
(which usually is more complicated). Comparing the results of the
original and the block action at a point in their respective coupling
constant spaces where the dimensionless correlation lengths agree, the
block action will definitely have better continuum properties, it has
less corrections to the leading scaling behaviour. It is ``closer to
the continuum limit''.

In an asymptotically free theory that issue can be resolved by
comparing the behaviour of the $\beta$-function in the weak coupling
region \cite{HaHa80,LuWe87,Ha89}. In our case in principle we can rely
on  renormalized perturbation theory.  In fact the proportionality
constant $C$ in the two-loop expansion for the renormalized mass in
\eq{ScalingMassRen} is regularization scheme dependent. We therefore
add a subscript $s$ denoting the regularization scheme (lattice action
or geometry). Asymptotically, at small enough small $\lambda_r$, $
a_s/a_{s^\prime} = C_s /C_{s^\prime}$ can be determined by lattice
perturbation theory.

For a comparison of different effective theories at finite cutoff and
not too small $\lambda_r$ we can no longer rely on that approach. It
would be appropriate to choose another physical observable $\cal O $,
that may serve as a means to measure the distance from the continuum
limit independently. Comparing results due to different actions $S_s$
(where the index distinguishes between the different actions and
regularization schemes) at the same values of $\cal O $ then provides
an estimate for the regularization dependence. This observable could be
for example a $\pi\pi$-scattering cross-section, the $\sigma$-decay
width or the violation of euclidean rotational invariance of some
quantities \cite{LuWe89,BhBiHe90,He90,GoKaNe93,LaWi93}. Different
operators $\cal O $ will usually have different cutoff dependence and
their choice needs further motivation: There is no ``best''
observable.  Quantities like the  ratio $R$ should be considered versus
such a ``distance'' $\cal O $ to the continuum limit.  Of course we
could still use the inverse correlation length $ m_{\mbox{\footnotesize phys}}
a_s$, but in order to compare different schemes we have to rescale
$a_s$ at each point such that $\mbox{$\cal O $}$ agrees.

In a Monte Carlo study of the 1-component $\Phi^4$ model \cite{LaWi93}
the violation of euclidean rotational invariance (in real space and in
momentum space) of the Higgs-propagator was used as measure for the
``distance to the continuum limit''. Four kinds of lattice actions,
respectively lattice geometries, were considered.
\begin{description}
\item[$\mbox{\boldmath $ C $}$ :] The standard action on the hypercubic
lattice \eq{HCaction}.
\item[$\mbox{\boldmath $ C' $}$ :] An action including diagonal nearest
neighbour terms with the same coupling \cite{BhBi88}.
\item[$\mbox{\boldmath $ F_4 $}$ :] The nearest
neighbour action on an $F_4$-lattice \cite{BhBiHe90,BhBiHe91}.
It's embedding into a hypercubic
lattice may be imagined by removing all odd sites of the grid.
The symmetries for this case
forbid terms like $\sum_\mu p_\mu^4$ in the inverse momentum space
propagator and guarantee Lorentz-invariance to a higher order
in the momentum cutoff $O(\Lambda^{-4})$ than the first two
hypercubic lattice schemes which had violating terms $O(\Lambda^{-2})$.
\item[$\mbox{\boldmath $ S $}$ :] The
Symanzik improved action
which  includes anti-ferromagnetic couplings to the next-to-nearest
neighbours along the 4 euclidean axis-directions.
These additional terms are incorporated in order to remove the
undesired $O(\Lambda^{-2})$ contributions. The remaining corrections
of $O(\Lambda^{-4})$ in the lattice dispersion
relation guarantee tree-level improvement \cite{Sy83}.
\end{description}
All four lattices may be embedded into the regular hypercubic one.

\FIG{RotInv}{RotInv.eps}{Fig. from \protect{\cite{LaWi93}}, showing the
amount of rotational symmetry in the inverse boson propagator in momentum
space, determined for 4 different actions.}{9cm}{9cm}{500}

Fig. \ref{RotInv} compares the rotational invariance properties for
the boson propagator in momentum space. The deviation from the
exact continuum symmetry may be quantified by considered the
coefficients in an expansion in spherical harmonics. Fig. \ref{RatioRotInv}
exhibits the final results for $R=m_\sigma/F$ (proportional to
the square root of the renormalized coupling) as functions of a measure
for the deviation from rotational symmetry. As discussed above,
this may be a more
appropriate measure than the correlation length itself. One find a
sizable difference, about 30-40\%, between the values for the actions
$\mbox{\boldmath $ C $}$ and $\mbox{\boldmath $ C' $}$ (which violate
rotational invariance to a larger extent) on one hand and
$\mbox{\boldmath $ F_4 $}$ and $\mbox{\boldmath $ S $}$  on the other
(which have no $O(p^4)$ violation in the tree-level propagator).

\FIG{RatioRotInv}{RatioRotInv.eps}{The ratio \protect{\eq{EqHiggsBound}}
vs. a measure of rotational invariance for the one-component model;
the curves are fits to
the asymptotic behaviour; the left-hand part give the results
on an extended scale. The symbols used denote the different
action ($C$: bars, $C'$: squares, $ F_4$: triangles, $S$: asterisks)
(fig. from \protect{\cite{LaWi93}}).}{12cm}{6.5cm}{380}

In the results for the $O(4)$ model
for $\mbox{\boldmath $ F_4 $}$ \cite{BhBiHe90} and
$\mbox{\boldmath $ S $}$
\cite{GoKaNe93} the authors estimated the
regularization dependence by comparing the results at given
value of another measure of rotational symmetry, derived from
a perturbation expansion of the Goldstone-Goldstone scattering
cross-section. There the difference between $\mbox{\boldmath $ F_4 $}$,
$\mbox{\boldmath $ S $}$ compared to
$\mbox{\boldmath $ C $}$ was smaller than that observed in \cite{LaWi93},
of the order of 10-20 \% for 1-10\%
violation of rotational symmetry. Although this variation could be due to the
study of different models (one component vs. four component
fields) it may also indicate a strong dependence on the choice of
measure for rotational invariance. Maybe systematic studies of higher order
terms, like
those initiated recently \cite{HeNeVr92,HeKlNe92,HeKlNe93a,JaKuLi93,JaKuLi93a}
are
helpful to clarify this issue.

Let us conclude this chapter with a summary of the Monte Carlo results
for the $O(4)$ $\Phi^4$ model.
\begin{itemize}
\item The observed scaling behaviour is consistent with the triviality
conjecture.
\item One clearly identifies Goldstone bosons and the massive Higgs
mode.
\item One finds good agreement with analytically derived results in
regions, where the comparison is possible ($a m_H<0.5$).
\item Finite size scaling works.
\item Chiral perturbation theory provides an excellent description of
$j>0$ finite volume results and allows a reliable extrapolation to
the thermodynamic situation of spontaneous symmetry breaking.
\item Allowing for a cutoff $>2 m_H$ (or ``continuum-like behaviour''
up to a few percent) one obtains an upper bound for the renormalized
coupling, that translates (using the experimental values for the gauge
coupling and the $W$-mass) into the bound on the Higgs mass \cite{He93},
\be
m_H < 710(60) \mbox{GeV} .
\ee
\item Some (relatively weak) regularization dependence of this bound
(and other quantities) has been observed.
\item Further work includes fermions (Yukawa models), but no drastic
change of the bound has been found.
\end{itemize}

The recent review of Heller \cite{He93} summarizes the situation and
gives further relevant references.

\section{Scattering: Phase Shifts and Resonances}
\setcounter{equation}{0}

Most observables that have been considered in lattice calculations so
far are {\em masses\/} (from propagators) and {\em coupling constants\/}.
Most experimental measurements in high energy physics, however, lead to
cross sections and their analyses to scattering phase shifts.  Indeed,
resonances are described by the slope and position of the values of the
phase shift passing through $\pi/2$ and the width is a quantity derived
under certain additional assumptions (like the Breit-Wigner shape). In
fact, considering the $O(4)$ $\Phi^4$ model we are in a situation,
where the Higgs boson is a resonance (coupling to Goldstone states)
and, strictly speaking, no asymptotic state of the theory. For the
determination of the Higgs mass it is therefore not sufficient to just
determine the single particle propagator over larger and larger
distances. In that limit one will only learn about the lowest lying
energy level in that channel, which is not the Higgs mass but a
2-particle threshold.

In finite systems the energy spectrum is discrete and may be related to
the phase shifts in the infinite volume system
\cite{Lu86a,Lu89,LuWo90,Lu91,Lu91a}. The leading volume dependence is
connected to the scattering lengths and early calculations in the 4D
Ising model determined the s-wave scattering length
\cite{MoWe87,FrJaJe90}. Meanwhile there are several studies:

\begin{itemize}
\item In a pioneering work L\"uscher and Wolff obtained phase shifts
in the 2D $O(3)$ nonlinear $\sigma$-model \cite{LuWo90}. There the result
is known analytically from conformal field theory and one found
agreement with the non-perturbative calculations.
\item Below I discuss another 2D study \cite{GaLa92,GaLa93,GaHiLa93}
with the aim to determine phase shifts in a resonating situation and to
find the resonance parameters in a non-perturbative way.
\item Fiebig et al. \cite{FiWo92,FiWoDo92} investigate the 3D QED and
determined
phase shifts for the scattering of mesonic bound states (in the
quenched situation, i.e. without dynamical fermions in the vacuum).
\item In the 4D  $O(4)$ $\Phi^4$ model there have been studies in the
symmetric phase \cite{Ni92} and in the broken phase with an additional
external symmetry breaking field \cite{Zi93,ZiWeGo93,WeZiGo93}. The latter
model is a test case for resonating phase shifts as well, since one has
a situation with light particles (the Goldstone bosons have non-zero mass
due to the external field) and one heavy resonance (the Higgs particle
decays into the Goldstone states and is no asymptotic state of the theory).
\end{itemize}
Only the efficient algorithms available for bosonic systems allow the
high precision that is necessary to determine the energy levels and
phase shifts, as discussed below. For this reason the scalar field
theories are the leaders in that approach -- the reason to discuss
approach and result in this last section of this review. Also, in that
process a lot can be learned on the representation of physical states
an the lattice.

\subsection{Energy Eigenspectrum of Finite Systems}

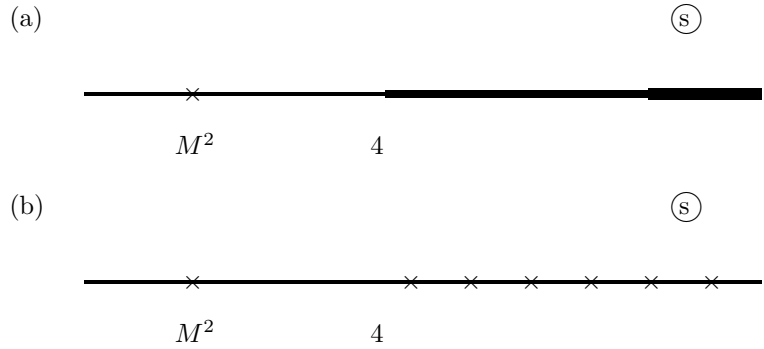
\begin{figure}
\begin{center}

\unitlength=0.1mm
\linethickness{1pt}
\begin{picture}(1300.00,600.00)
\put(200.00,450.00){\line(1,0){900.00}}
\put(600.00,453.00){\line(1,0){500.00}}
\put(600.00,447.00){\line(1,0){500.00}}
\put(950.00,456.0){\line(1,0){150.00}}
\put(950.00,444.0){\line(1,0){150.00}}
\put(330.0,440.0){$\times$}
\put(320.00,370.00){$M^2$}
\put(580.00,370.00){4}
\put(100.0,540.0){(a)}
\put(990.00,540.00){s}
\put(1000.00,550.00){\circle{40}}
\put(200.00,200.00){\line(1,0){900.00}}
\put(330.0,190.0){$\times$}
\put(620.0,190.0){$\times$}
\put(700.0,190.0){$\times$}
\put(780.0,190.0){$\times$}
\put(860.0,190.0){$\times$}
\put(940.0,190.0){$\times$}
\put(1020.0,190.0){$\times$}
\put(320.00,120.00){$M^2$}
\put(580.00,120.00){4}
\put(100.0,290.0){(b)}
\put(990.00,290.00){s}
\put(1000.00,300.00){\circle{40}}
\end{picture}
\vspace{-1cm}
\end{center}
\caption{The spectral density in the $s=W^2$-plane;
(a) infinite volume, (b) finite volume.\label{AnalyticSPlane}}
\end{figure}

Consider the propagation of a state $| N \rangle$ in a channel with
1-particle (mass $M$) and 2-particle states (mass $m=1$, i.e. we choose
the light state as the mass unit for simplicity of notation, where not
stated otherwise). In infinite volume the analytical structure of the
propagator in the relativistic variable $s$ is as shown in fig.
\ref{AnalyticSPlane}(a), with an elastic cut starting at $4 m^2\equiv
4$, higher inelastic thresholds and (for $M<2m$) a bound state below
the elastic threshold. The dispersion relation representation gives
\be
T(s) = \mbox{pole} + \frac{1}{\pi} \int_4^\infty ds'{\rho(s')\over (s'-s)}
\ee
or, Fourier transformed to euclidean time $t$,
\be
G(t) = c_0 e^{-Mt} + \frac{1}{\pi} \int_4^\infty dW'e^{-W't} \rho(W')
\ee
where $W$ denotes the energy and $\rho$ is the spectral density. If the
single particle state is above threshold it describes a resonance and
is no longer part of the discrete spectrum and no longer an asymptotic
state.

In finite volume the continuous spectrum is replaced by a
discrete set (cf. fig. \ref{AnalyticSPlane}(b)),
\be
G(t) = c_0  e^{-Mt} + \sum_i c_i e^{-W_i t} ,
\ee
with the eigenspectrum $\{W_i\}$. A possible resonance is represented
by the form of the discrete spectrum and not by
a single energy level like a bound state.

How can we determine the eigenspectrum in realistic lattice
calculations? The answer is given by standard methods of quantum
mechanics. One has to find a (sufficiently) complete set of
operators $N_n$ with the correct quantum numbers
(symmetry properties) of that channel. Each $N_n(t),   n = 1,2...$ is
defined on a single time slice, and $t$ denotes the separation of the
time slices. In the Monte Carlo simulation one then measures the
connected correlator,
\be\label{OpCorrFunc}
C_{nm}(t) = \langle {N_n}^*(0) N_m (t) \rangle_c \equiv
\langle {N_n}^*(0) N_m (t) \rangle -
\langle {N_n}^* \rangle \langle N_m \rangle \; .
\ee
The transfer matrix formalism (see e.g. \cite{ItDr89}) yields
the spectral decomposition
\bea\label{Spectraldecotrunc}
C_{nm}(t) &=&\sum_l  \langle N_n^*| l \rangle e^{-W_l t}\langle l |
N_m\rangle_c\\
&=&\sum_l v_n^{(l)*}v_m^{(l)} e^{-W_l t}\; .
\eea
For simplicity we assume non-degenerate $W_l$, ordered
increasingly. The amplitudes $v^{(l)}_n = \langle l | N_n \rangle$
are the projections of the states $| N_n \rangle$ (generated by the
operators $N_n$ out of the vacuum) on the energy eigenstates
$\langle l|$ of the scattering problem.

In actual calculations one has to truncate the correlation matrix,
since one can consider only a finite number $r$  of operators. The
generalized eigenvalue problem
\be \label{Geneigprob}
C( t ) \; {\ZETA}^{(k)}(t) = \lambda^{(k)} ( t , t_0 ) \;
C( t_0 ) \;
{\ZETA}^{(k)}(t) \QA k = 1,2, .. , r \;.
\ee
allows one to find the energy eigenvalues $W_l$ efficiently even for
comparatively small values of $t$.
In \eq{Geneigprob} we assume $t_0<t$, e.g. $t_0=1$.
We require the $N_n \; , \; n = 1,2,..,r$ to be linearly independent,
thus $C( t_0 )$ is regular and the generalized eigenvalue problem is
well defined. Indeed, it may be transformed to a standard eigenvalue
problem for the matrix
\be \label{matrixT}
C^{-\ot} ( t_0 ) \; C ( t ) \; C^{-\ot} ( t_0 ),
\ee
with identical eigenvalues $\lambda^{(k)} ( t , t_0 )$
and eigenvectors
$\vecu^{(k)} (t) = C^{\ot} ( t_0 ) {\ZETA}^{(k)}(t)$.
The solution for the eigenvalues is
\be \label{Eigenvalue}
\lambda^{(k)} ( t , t_0 ) = e^{- ( t - t_0 ) W_k} \; .
\ee

Some approximations have to be done in the numerical calculations,
among them the restriction to a finite set of lattice operators and
therefore to a finite set of eigenvalues.  However, as has been pointed
out in \cite{LuWo90}, the correction due to the truncation is of
$O(\exp [ -(t-t_0) W_{r+1} ])$ by a perturbation calculation. The range
of $t$-values is dictated by statistical errors of the values of the
correlation function and is usually between 1 and 8.

\subsection{Spectrum and Phase Shifts}

Consider quantum mechanics in a $(d-1)$-dimensional spatial box, with
periodic (torus) boundary conditions with cubic symmetry.  Assume that
the interaction region is localized and smaller than the extension of
the box.  Solving the static Schr\"odinger equation outside the
interaction region ($V=0$, i.e. the Helmholtz equation) and expanding
the solution into cubic ($SO(d-1,\Ga )$) spherical harmonics one finds a
relation between the discrete values of the momenta and the phase shift
at these points \cite{Lu89,LuWo90,Lu91}. As a simple example we
consider the scalar $A_1^+$ sector and neglect higher angular momenta
($>3$). We find
\bea
\exp(2i\delta_0(k)) &=& {m_{00}(k)+i \over m_{00}(k)-i}\; ,\\
m_{00}(k)&=&{\cal Z}_{00}(1;q^2) / q \pi^\frac{3}{2} \quad\mbox{where}\quad q
= {kL\over 2\pi}\; ,
\eea
i.e. $m_{00}(k)=\cot \delta_0(k)$, the usual $K$-matrix. The function
${\cal Z}_{00}$ is a generalized zeta-function, and one has
\be
\begin{array}{rcll}\label{DefofPhi}
m_{00}(k)&=&\cot q\pi &\mbox{for d=2}\; ,\\
&=&\displaystyle  {1\over 4 \pi^2q}\sum_{\vec{n}\in Z^3} {1\over \vec{n}^2-q^2}
&\mbox{for d=4}\; .
\end{array}
\ee
This gives a relation
\be \label{DeltavsPhi}
\delta(k_n)=n\pi - \Phi (k_n)
\ee
between the quantized values of $k$ and the phase shift. For $d=4$ the
functional form of $\Phi(k)$ is determined numerically (cf.
\cite{ZiWeGo93}).
For $d=2$ one has the simple form $\Phi (k_n)=k_n L/2$ and thus
\be
2\delta(k_n) + k_n L = 2 n \pi \quad (n\in\Ga) \; ,
\ee
and this quantization condition may be easily visualized for the cases
without and with interaction (fig. \ref{PhaseQuantization}). Because of
the periodic b.c. of the wave functions one has $\Psi(0)=\Psi(L)$,
$\Psi(0)=\Psi(L)$. Due to the (localized) interaction a non-zero phase
shift affects the quantization of the momenta.

\FIG{PhaseQuantization}{PhaseQuantization.eps}{The quantization in d=1
(a) without and (b) with interaction phase shift.}{8.5cm}{4cm}{600}

These arguments and the equation \eq{DefofPhi} may be applied to a
quantum field theory problem under the following restrictions and
assumptions.
\begin{itemize}
\item The interaction region is localized $<L$ and, in
particular, the size of the single particle states is smaller than the
lattice: $1/m\ll L$. This is due to that one wants a pair of two free
particles as asymptotic states.
\item The polarization effects are
controllable. Of course there will be interaction, e.g. self
interaction of the light particles around the torus world. In fact this
is perturbatively the leading contribution. This effect can be
understood and accounted for.
\item On is confined to the elastic
regime $4<s<9$ (for a 3-particle inelastic threshold, or 16 for a
4-particle threshold, depending on the quantum number of the channel).
\item One understands the lattice artifacts (they are $O(a^2)$). This
is because we work close to but never at the phase transition. These
terms are due to the grained structure of the lattice and leading
contributions may be studied already in the non-interacting gaussian
model.
\end{itemize}

For the determination of the energy spectrum one should consider
correlation functions of a sufficiently large number of observables
with the correct quantum numbers, capable to represent the eigenspace of
scattering states \cite{LuWo90}.

To summarize, for a given set of couplings the determination of the
phase shifts in the scattering sector (at total momentum zero) consists
of the following steps.
\begin{enumerate}
\item For given lattice size $L$ determine the single particle state;
check for polarization effects and get the mass $m$.
\item Determine the correlation matrix $C ( t )$ for various values of
$t$, solve the eigenvalue problem for this matrix, and obtain the
spectrum from the exponential decay \eq{Eigenvalue}. The dimension $r$
of $C$ cannot  be chosen too large, because then the inversion
becomes numerically unstable. We denote the energy eigenvalues by
$W_n(L)$.
\item From the values $W_n(L)$ one finds $k_n$ with help of the
dispersion relation
\be\label{Wdispersionrelation}
W_n=2\sqrt{m^2+k_n^2} .
\ee
This uses the fact that the 2-particle energy is the sum of the (back
to back) outgoing single particle energies.
\item From the values $k_n$ one obtains $\delta(k_n)$ from
\eq{DeltavsPhi}.
\end{enumerate}
For each lattice size $L$ one obtains in this way several (typically
the lowest 1--4) energy eigenvalues and thus pairs $k_n$,
$\delta(k_n)$. The procedure is then repeated for other
lattice sizes. Eventually one gets more and more values $k$ and $\delta$,
filling the domain $0<k<\sqrt{5/4}$ (or $\sqrt{3}$ if the 4-particle
threshold is the first inelastic one).
In fig. \ref{FreeSpectrum} the spectrum of free 2-particle states
(vanishing phase shift) for the 4D system is shown.

\FIG{FreeSpectrum}{FreeSpectrum.eps}{The free 2-particle energy spectrum
($\delta=0$) according to \protect{\eq{DeltavsPhi}}; the density increases
with the volume (from \protect{\cite{Zi93}}).}{8.5cm}{6.5cm}{400}

\subsection{Models and Results}

Here I want to discuss two studies of scalar models
and their results, concentrating on common features.

\subsubsection{2D Resonance Model}
The authors \cite{GaLa92,GaLa93,GaHiLa93} study a model, which
describes two light particles $\varphi$ that
couple to a heavier particle $\eta$ giving rise to resonating behaviour.
For this one couples two Ising fields through a 3-point term,
\be
S = S_{\mbox{\footnotesize Ising}}(\varphi,\kappa_\varphi)
    + S_{\mbox{\footnotesize Ising}}(\eta,\kappa_\eta)
    +\frac{g}{2}\sum_{x \in \Lambda, \mu=1,2} \eta_x \varphi_x
            (\varphi_{x-\hat{\mu}} + \varphi_{x+\hat{\mu}})   \; .
\ee
The values of the fields are restricted to $\{ +1, -1\}$ and
$S_{\mbox{\footnotesize Ising}}$ denotes the usual action of the Ising model.
The 3-point term is introduced in a nonlocal manner,
because $ \varphi_x^2 \equiv 1$. The system is studied in the symmetric
phase with regard to both nearest-neighbour coupling $\kappa_\varphi$
and $\kappa_\eta$.

The case $g = 0$ corresponds to the situation of two independent Ising
models, each with a 2nd order phase transition at $\kappa_c =\ot \ln
(1+\sqrt{2}) \simeq 0.44068$.  In the scaling limit the model describes
an interacting boson with mass $m = - \log ( \tanh \kappa) - 2 \kappa$
\cite{ItDr89}.  Reformulating  the Ising model at the phase transition
as a theory of non-interacting fermions, it was shown that the
scattering matrix assumes the value $- 1$ independent of the momentum
\cite{SaMiJi77}.

In the coupled case ($g > 0$), if we identify $\eta$ and $\varphi$ with
particle states, the term proportional to $g$ gives rise to transitions
like $\eta \rightarrow \varphi \varphi$ rendering $\eta$ a resonance in
the $\varphi\varphi$ channel, when kinematically allowed.  The phase
diagram of this model has interesting structure with limiting cases
including random external fields and random bond models.  Its structure
is discussed in more detail in \cite{GaLa93}.

The values of the couplings $\kappa_\varphi, \kappa_\eta$ are adjusted
such that $m_\varphi=0.19$ and $m_\eta=0.50$. The system is studied for
the case of two independent Ising models $g=0$ as well as for the
values $g$=0.02, 0.04. The lattice sizes $L\times T$ considered were in
the range $L=12\ldots 60$ and $T=60\ldots 100$. Cluster updating was
used, with typically 200 K measurements per point. In the 2-particle
channel 4--6 operators $N_n$ were considered and 2--5 lowest
eigenvalues determined from diagonalizing the correlation function at
distances $\Delta t =2\ldots 8$.

\subsubsection{$O(4)\;\Phi^4$ Model}
Zimmermann et al. \cite{Zi93,ZiWeGo93,WeZiGo93} study the
$\lambda\to\infty$ limit of the $O(4)$ model, i.e. the nonlinear
$\sigma$-model, with a small external field. As discussed, this is a
theory with three degenerate light pions $\vec{\pi}$
(would-be-Goldstone bosons) and a heavy $\sigma$ (the Higgs boson).

The bare coupling parameters $\kappa$ and $j$
were adjusted to have renormalized masses $m_\pi=0.23$ and $m_\sigma=0.70$.
The system was simulated with the cluster algorithm on lattices of size
$L^3\times T$ with $L=12\ldots32$ and $T=32\ldots 40$ (40-50 K
measurements per point). The authors considered 6 operators in the
2-particle channel and determined typically 2--5 lowest energy
eigenvalues from the correlator at distances $\Delta t =2\ldots 8$.

\subsubsection{Single Particle States}
For the determination of the energy spectrum a precise knowledge of the
single particle mass and related finite size effects is important.
The operator of a $\varphi$ state with momentum
\be
p_{\nu} = 2 \pi \nu /L \QA \nu = -L/2+1, \ldots, L/2
\ee
may be represented through
\be
\label{Operatorphi}
\varphi(p_\nu,t)= \frac{1}{L^{d-1}}\sum_x
\varphi(x,t) \exp{(i p_\nu x)} \;.
\ee
Its connected correlation function over temporal distance $t$ decays
exponentially $\propto \exp{(-E_\nu |t|)}$ defining the single particle
energy $E_\nu$; in particular we have $E_{\nu=0} = m_\varphi$.

The observed mass, as compared to the ``real'' mass at vanishing
lattice spacing and infinite volume, incorporates contributions from
polarization due to self interaction around the torus
(decreasing exponentially with L)
and lattice artifacts due to the finite ultra-violet cutoff
(polynomial in the lattice constant a).

\FIG{SinglePartProp}{SinglePartProp.eps}{Results ($E_\nu, p_\nu)$ as
obtained from the analysis of the single particle propagator in the
2D Ising model ($g=0$, $L=50$, from
\protect{\cite{GaLa93}}). The dashed curve represents the continuum d.r.
\protect{\eq{ContDR}} and the full curve the lattice d.r.
\protect{\eq{latticeDR}}.}
{8.5cm}{6.5cm}{500}

Fig. \ref{SinglePartProp} is a plot of $E_\nu$ vs. $p_\nu$,
obtained for the pure
Ising model ($g=0$) and $L=50$. Comparing the values ($E_\nu, p_\nu)$
with the continuum spectral relation ,
\be\label{ContDR}
E_\nu = \sqrt{m^2 +p_\nu^2}
\ee
one observes deviations $O( (a p)^2)$, as expected \cite{LuWo90}.
Using the dispersion relation for
the lattice propagator of a gaussian particle with mass $m$,
\be\label{latticeDR}
E_\nu= \mbox{arcosh}( 1-  \cos p_\nu+ \cosh m )\; ,
\ee
one finds excellent agreement with the measured values. Thus
this particular single particle state shows little deviation from the
free lattice form. These results for the 2D Ising model holds for the
situations $g\neq 0$ as well and was confirmed in the 4D
study \cite{Zi93,ZiWeGo93,WeZiGo93}, too.

Let us, for the moment, stay with the 2D Ising model.  It may be
considered the limit of a $\Phi^4$ theory with infinite bare 4-point
coupling $\lambda$.  We are in the symmetric phase, thus we expect the
leading interaction to be due to an effective 4-point term.  The
finiteness of the spatial volume allows self-interaction with one
particle running around the torus.  This leads to a finite volume
correction to the particle mass proportional to $\lambda_r e^{- m L}
/\sqrt{L}$.  In the 2D model excellent agreement with this behaviour
was observed. This mass shift on finite lattices has also been
confirmed in Monte Carlo simulations of the Ising model in 4 dimensions
\cite{MoWe87}.

\subsubsection{Scattering Sector}
In the 2-particle channel one considers operators
with total momentum zero and quantum numbers of the $\eta$,
a set as complete as possible. These operators may be constructed
from
the $p=0$ single $\eta$-operator,
\be
N_1(t)\equiv \eta(p=0,t) = {1 \over L^{d-1}} \sum_x \eta(x,t)
\ee
as well as from the combination of two $\varphi$-operators with
opposite momentum,
\bea\label{Operators}
N_n(t) &=& \sum_p f_n(p) \varphi(-p,t) \varphi(p,t)\; ,\\
\varphi(p,t) &=& {1 \over L^{d-1}} \sum_x
e^{2i\pi (x.p)/L } \varphi(x,t) \; ,
\eea
where $p$ and $x$ have $d-1$ (integer valued) components.
The wave function in
momentum space $f_n(p)$ has been chosen in various ways.
\begin{itemize}
\item
Plane waves would be the eigenstates for the non-interacting
case. This corresponds to choosing all integer values of
$|p|^2$,
\be
\begin{array}{rclcl}\label{planewaves}
f_n(p) &=& \delta(n,|p|)   &\mbox{for}&d=2 \;,\\
f_n(p) &=& \delta(n,|p|^2) &\mbox{for}&d=4 \; .
\end{array}
\ee
\item
However, one could also choose special forms like the periodic
singular solutions of the Helmholtz equation
\be
f_n(p)={ 1\over p^2 - q_n^2}
\ee
as has been tried in \cite{ZiWeGo93}.
\end{itemize}
The question here is which choice gives the best representation of the
scattering eigenstates. Any ``complete'' set would do in principle, but
in practice one is limited to a small number of operators and thus
there may be ``bad'' and ``better'' choices. In the calculations
discussed here no clear priority for one of the discussed forms was
found. Both worked equally good; thus one favours the simpler
form \eq{planewaves}.

These operators and products thereof are now measured for each
configuration and the resulting correlation function \eq{OpCorrFunc} is
computed for various values of $t$, diagonalized and the energy
spectrum determined
from the exponential decay \eq{Eigenvalue}.

\FIG{EnergySpec12}{EnergySpec12.eps} {The lowest energy levels for (a)
the non-interacting gaussian model and (b) the Ising model; the
decoupled $\eta$-state is also shown. The full lines denote curves due to
phase shifts (a) $\delta=0$  and (b) $\delta=-\frac{\pi}{2}$
(fig. from \protect{\cite{GaLa93}}).}{11cm}{6cm}{400}

\FIG{IsingDelta}{IsingDelta.eps} {The phase shift obtained for the
Ising model from the energy spectrum in fig.
\protect{\ref{EnergySpec12}} (fig. from \protect{\cite{GaLa93}}).}
{7.7cm}{5cm}{500}

In fig. \ref{EnergySpec12} we compare the results for the energy levels
for the 2D Ising with those for the free, non-interacting case. We find
that the lowest level moves from the threshold (in the free case) to
higher values in the Ising situation. In the figure we also plot the
results of the (decoupled) single $\eta$ state at $m_\eta=0.5$. The
level crossing therefore is fictitious and does not really occur, since
the $\varphi\varphi$ channel is decoupled from the $\eta$-channel
($g=0$).  We note that we have only a few
energy eigenstates between the elastic threshold
(at 0.38) and the inelastic threshold (at 0.76).

The critical 2D Ising model has an $S$-matrix equal to $-1$
and we therefore expect a phase shift $\dising =
-\frac{\pi}{2}\pmod{\pi}$ in the scaling regime. Fig. \ref{IsingDelta}
shows the results for the phase shift $\delta (k)$ determined from the
data for the Ising model (fig. \ref{EnergySpec12}(b)) as a function of
the dimensionless momentum $k/m_\varphi$.

Figs. \ref{EnergyPhase34} exhibit results of the 2D resonance model in the
coupled situation, for $g\neq 0$. For most of the points the errors are
smaller than the symbols, typically $0.5\%$ of the energy value. The
degeneracies in the energy spectrum have now disappeared, resulting in
avoided level crossing with a gap that grows with $g$.

\FIG{EnergyPhase34}{EnergyPhase34.eps} {Left: The measured energy
levels for the $\varphi\varphi$ channel for the coupling constant
values $g=0.02$, vs. the spatial lattice extension $L$. The full curves
show the theoretical expectations for the values from the phase shift
as discussed in the text. The dashed lines indicate the 2- and the
4-particle thresholds.  Right: The phase shift as determined from the
energy levels (fig. from \protect{\cite{GaLa93}}).}{13cm}{6.2cm}{420}

One should keep in mind that points at neighboured values of $k$ may
come from quite different values of $L$ and different branches of
energy levels. In order to emphasize this situation corresponding
symbols denote the energy values on one branch, i.e. for one value of
$n$. Each branch, from small to large $L$ contributes to the whole
range of  $k$ values; it starts at larger values of $W$ (or $k$) with a
phase shift close to $\pi/2$, passes through the resonance value at the
plateau and approaches $-\pi/2$ towards smaller $k$. The overall
consistent behaviour is impressive.  The high energy values,
corresponding to large $k$ may feel the lattice cutoff effects $O(a^2)$
as well as possible misrepresentation of the energy eigenstates by the
considered operators. Within the accuracy of the data none of these
effects seems to be a problem.

The observed overall behaviour has a simple interpretation:  A
resonating phase shift superimposes the background Ising phase shift.  The
resonance parameters may be determined from the phase shift like from
experimental data, e.g. by a $K$-matrix fit or an effective range
approximation.

\FIG{EnergyLevels4D}{EnergyLevels4D.eps} {The energy levels in both
s-wave channels of the 4D $O(4)$ model for a particular choice of
couplings ($m_\pi=0.295, m_\sigma= 0.906$);
the dashed curves denote results due to renormalized
perturbation theory (from \protect{\cite{Zi93}}).} {12.5cm}{6.5cm}{320}

\FIG{O4PhaseShifts}{O4PhaseShifts.eps}{The isospin 0 s-wave phase shift
with the Higgs resonance determined from the energy spectrum in a
non-perturbative Monte Carlo simulation (from \protect{\cite{Zi93}}).}
{8.7cm}{6.5cm}{400}

Similar results were obtained in the study of the 4D $O(4)$ model.  In
4 dimensions the numerical task is much more demanding and therefore
one could not study as broad a range of lattice sizes as in the 2D
case.  However, the region of the first avoided level crossing could be
reached, as demonstrated in fig. \ref{EnergyLevels4D}. Fig.
\ref{O4PhaseShifts} gives the isospin 0 s-wave phase shift; this is the
channel with the Higgs resonance.

\subsection{What Can We Learn?}

\subsubsection{Comparison to Perturbation Theory}
The value of the partial width agrees with the Born term approximation
using the value of the renormalized 3-point coupling
\cite{Zi93,WeZiGo93}.  This implies that the situation is indeed
described by an effective Lagrangian with tree-level expansion.
However, the effective couplings have to be determined from the
non-perturbative Monte Carlo calculation.

\subsubsection{Representation of Operators in the Scattering Channel}
In realistic situations we have to consider scattering between
boundstates: In QCD the
asymptotic states are bound states of the original quark fields,
indeed a very demanding situation. One of the results of the
purely bosonic scattering studies presented above concerns the quality
of the representation of states through the lattice operators $N_n(t)$.

If the set of lattice operators $N_n(t)$ is complete, we may relate the
eigenvectors of the diagonalization \eq{matrixT}
to the physical eigenvectors
${\ZETA}^{(k)}(t)\propto \vecv^{(l)}$ \cite{GaLa92}.
Let us express the original approximation \eq{Spectraldecotrunc}
to $C(t)$ in terms of these eigenvectors,
\be \label{NewSpectraldecotrunc1}
C_{n m} ( t ) = \sum_{l=1}^{r} {\zeta^{(l)}_n}^* \zeta^{(l)}_m
| \vecv^{(l)}| ^2 e^{-t W_l} \QA n,m = 1,2\ldots r
\; .
\ee
In particular consider the diagonal element
\be \label{NewSpectraldecotrunc2}
C_{nn} ( t ) = \sum_{l=1}^{r} | \zeta^{(l)}_n| ^2
  | \vecv^{(l)}|^2 e^{-t W_l} \; .
\ee
For a continuous energy spectrum,
\be \label{ContSpectraldecotrunc}
C_{nn}(t) = \int dW \rho_n(W) e^{-t W} \; ,
\ee
and we identify  $|\zeta^{(l)}_n| ^2 |\vecv^{(l)}|^2
\propto \rho_n(W_l)$, the spectral density of the
correlation function of operator $N_n$.
We therefore find that $|\zeta^{(l)}_n| ^2$ indicates the
relative weight of the contribution of operator $N_n$ to the energy
eigenstate $|l \rangle$.

\FIG{OpRepr}{OpRepr.eps} {Contributions of the operators $N_1 , N_2$
and $N_3$ (circles, triangles, squares) to the energy eigenstates
(a) $|1\rangle$ , (b) $|2\rangle$ and  (c) $|3\rangle$,
displayed as a function of the lattice size $L$ (fig. from
\protect{\cite{GaLa93}}).}{11cm}{4.5cm}{400}

Fig. \ref{OpRepr} shows the relative weights $| \zeta^{(l)}_n|
^2$ for $l=1,2,3$ and $n=1,2,3$ for the 2D resonance model (case
g=0.04). From the figure it can be seen, that the most important
contributions to an eigenstate come from lattice operators that have
similar eigenenergies in their eigensystem.  Compare the energy
spectrum of fig.s \ref{EnergySpec12} and \ref{EnergyPhase34} with fig.
\ref{OpRepr}.  For small $L$ the lowest energy state $W_1$ is close to
the resonance energy. There the state is completely dominated by the
$\eta$-operator $N_1$ (circles in fig.  \ref{EnergyPhase34}), which
describes an $\eta$-particle at rest.  For increasing $L$, $W_1$
decreases and approaches the 2-particle threshold. This manifests
itself in a drastic decline of the contribution of $N_1$, accompanied
by an increase of the amplitudes for two $\varphi$-particles at rest
(operator $N_2$, triangles) and with relative unit-momentum (operator
$N_3$, squares).  For lattice size larger than 20, the energy $W_1$ is
already too far below the resonance energy so that the energetically
higher $\eta$ at rest cannot contribute much to $| 1\rangle $.

In this way we may discuss the other levels as well. E.g.  the
contribution of the $\eta$-operator $N_1$ (circles) to the second
lowest state $|2\rangle $ assumes peak values where it crosses the
resonance energy plateau.

Also, with this consideration we can identify misrepresentation of
energy eigenstates  in terms of the considered operators. At higher
energies one observes a shift of the weight  factors
$| \zeta^{(l)}_n| ^2$ towards higher operators $N_n$.
If one has a situation, where the center of weight
lies outside the scope of operators considered, more operators should
be included for a good representation of that state \cite{GaLa92}.  The
study of the representation of scattering states by the operators
entering the correlation functions therefore sheds light on the problem
of constructing good approximations for the energy eigenstates
and it allows to control unwanted truncation effects.

\subsubsection{Lattice Artifacts}
In \cite{GaHiLa93} the calculations for the Ising model were repeated for
different coupling, corresponding to coarser lattice spacing.
One found a stronger deviation
from the expected phase shift at $a m_\varphi=0.5$ than at $a
m_\varphi=0.19$, compatible with the expected corrections $O(a^2)$.
However, as mentioned earlier, the d.r. \eq{Wdispersionrelation}
gives the total energy of the
asymptotic 2-particle state which (under the assumption of localized
interaction region) is just twice the energy of the outgoing particles.
Now the $O(a^2)$ corrections of the single particle d.r. have
been nicely described by replacing the continuum d.r. by the
lattice relation \eq{latticedispersionrelation}.
If one   replaces \eq{Wdispersionrelation} by the corresponding
lattice expression,
\be \label{latticedispersionrelation}
W_n = 2 \,\mbox{arcosh}(1 -\cos k_n + \cosh m )\; ,
\ee
the data for $W_n$ now produce slightly different values of $k_n$ and
$\delta$, in much better agreement with a constant value of
$-\pi/2$. One concludes, that the leading $O(a^2)$ corrections can be
expressed by replacing the continuum d.r. by the lattice relation.
This observation was confirmed in the study of the 4D model \cite{Zi93}.

In summary, it seems that L\"uscher's method to determine phase shifts
from the volume dependence of the discrete energy spectrum works nicely,
also for resonances. There are problems, where some of the
restrictions cannot be met. In cases where more than one channel
has non-vanishing phase shifts the determining equation involves
matrices and the identification becomes problematic. Most likely
one needs further continuity arguments like those used in the phase
shift analyses of experiments.

\section*{Conclusion}

During the last decade computer simulation have become an important
tool in QFT. Within the realm of purely bosonic field theory efficient
stochastic algorithms have been developed, that allow high precision
results. Almost always, however, it is insufficient to just let the
programs run and wait long enough until the answer stabilizes. Finite
size and relaxation problems are persistent. In certain limits we may
have an idea of the structure of an underlying (possibly: effective)
field theory that will have unknown parameters and a limited domain of
validity.  In this way different effective models may describe
different limiting cases of the microscopic theory.  The
nonperturbative Monte Carlo calculations can provide the unknown values
of the parameters in {\em ab initio\/} calculations.
This has made the results based on a finite volume theory (like finite
size scaling from renormalization group, chiral perturbation theory, or
the finite volume dependence in the situation of phase shifts, to
mention the examples discussed here) the most reliable ones.

The good agreement of Monte Carlo results with analytical expectations
has opened the way for more sophisticated applications like using the
Monte Carlo approach as an ``experimental'' test laboratory for
quantum field theory models.  Obviously one cannot hope to prove something
(maybe one can disprove some ideas) but one can hope to identify
relatively quickly promising or most-likely-bad theories.  An example
for this approach is the study of Yukawa-models and chiral fermions
(cf.  the review in \cite{DeJe93}). The more conservative reason for
Monte Carlo studies, and still the main motivation, is the possibility
to determine numbers in this direct, non-perturbative quantization
scheme, that are up to now not at all obtainable with other methods.


\begin{thebibliography}{100}

\bibitem{Ai81}
M. Aizenman,
\newblock Phys. Rev. Lett. {\bf 47} (1981) 1.

\bibitem{Ai82}
M. Aizenman,
\newblock Commun. Math. Phys. {\bf 86} (1982) 1.

\bibitem{AiGr83}
M. Aizenman and R. Graham,
\newblock Nucl. Phys. B {\bf 225 [FS9]} (1983) 261.

\bibitem{Ba83}
M.~N. Barber,
\newblock in {\em Phase Transitions and Critical Phenomena}, edited by C. Domb
  and J. Lebowitz, volume VIII, Academic Press, New York, 1983.

\bibitem{BeNe91}
B. Berg and T. Neuhaus,
\newblock Phys. Lett. B {\bf 267} (1991) 249.

\bibitem{BeNe92}
B. Berg and T. Neuhaus,
\newblock Phys. Rev. Lett. {\bf 68} (1992) 9.

\bibitem{BhBi88}
G. Bhanot and K. Bitar,
\newblock Phys. Rev. Lett. {\bf 61} (1988) 798.

\bibitem{BhBiHe90}
G. Bhanot, K. Bitar, U.~M. Heller, and H. Neuberger,
\newblock Nucl. Phys. B {\bf 343} (1990) 467.

\bibitem{BhBiHe91}
G. Bhanot, K. Bitar, U.~M. Heller, and H. Neuberger,
\newblock Nucl. Phys. B {\bf 353} (1991) 551.

\bibitem{Bi92}
K. Binder,
\newblock Finite size effects at phase transitions,
\newblock in {\em Computational Methods in Field Theory Lecture Notes in
  Physics 409}, edited by H. Gausterer and C.~B. Lang, page~59,
  Springer-Verlag, Berlin, Heidelberg, 1992.

\bibitem{BlSw80}
H.~W.~J. Bloete and R.~H. Swendsen,
\newblock Phys. Rev. B {\bf 22} (1980) 4481.

\bibitem{BoHaHa86}
K.~C. Bowler et~al.,
\newblock Phys. Lett. B {\bf 179} (1986) 375.

\bibitem{Br82}
E. Br{\'e}zin,
\newblock J. Physique {\bf 43} (1982) 15.

\bibitem{BrLeZi76}
E. Br{\'e}zin, J.~C. {Le~Guillou}, and J. Zinn-Justin,
\newblock Field theoretical approach to critical phenomena,
\newblock in {\em Phase Transitions and Critical Phenomena}, edited by C. Domb
  and M.~S. Green, volume~VI, page 127, Academic Press: New York, 1976), 1976.

\bibitem{BrFrSo83}
D. Brydges, J. Fr{\"o}hlich, and A.~D. Sokal,
\newblock Commun. Math. Phys. {\bf 91} (1983) 141.

\bibitem{BrFrSp82}
D.~C. Brydges, J. Fr{\"o}hlich, and T. Spencer,
\newblock Commun. Math. Phys. {\bf 83} (1982) 123.

\bibitem{BuLe82}
T.~W. Burkhardt and J.~M.~J. van Leeuwen,
\newblock Progress and problems in real-space renormalization,
\newblock in {\em Real-space Renormalization}, edited by T.~W. Burkhardt and
  J.~M.~J. van Leeuwen, page~1, Springer-Verlag Heidelberg, 1982.

\bibitem{BuLe82a}
T.~W. Burkhardt and J.~M.~J. van Leeuwen, editors,
\newblock {\em Real-space Renormalization},
\newblock Springer-Verlag Heidelberg, 1982.

\bibitem{CaMaPa79}
N. Cabibbo, L. Maiani, G. Parisi, and R. Petronzio,
\newblock Nucl. Phys. B {\bf 158} (1979) 295.

\bibitem{Ca88a}
D.~J.~E. Callaway,
\newblock Phys. Rep. {\bf 167} (1988) 241.

\bibitem{CaPe84a}
D.~J.~E. Callaway and R. Petronzio,
\newblock Nucl. Phys. B {\bf 240 [FS12]} (1984) 577.

\bibitem{Ca88}
J.~L. Cardy,
\newblock in {\em Finite-Size Scaling}, edited by J.~L. Cardy, page~1,
  North-Holland, Amsterdam, 1988.

\bibitem{CoKl80}
A. Coniglio and W. Klein,
\newblock J. Phys. A {\bf 13} (1980) 2775.

\bibitem{Cr83}
M. Creutz,
\newblock {\em Quarks, Gluons and Lattices},
\newblock Cambridge Univ. Press, 1983.

\bibitem{Cr92}
M. Creutz,
\newblock {\em Quantum Fields on the Computer},
\newblock World Scientific, Singapore, 1992.

\bibitem{DaNe83}
R. Dashen and H. Neuberger,
\newblock Phys. Rev. Lett. {\bf 50} (1983) 1897.

\bibitem{DeJe93}
A.~K. De and J. Jers{\'a}k,
\newblock Yukawa models on the lattice,
\newblock in {\em Heavy Flavours}, edited by A.~J. Buras and M. Lindner, page
  732, World Scientific, Singapore, 1993.

\bibitem{ArCaFr83}
C.~A. de~Carvalho, S. Caracciolo, and J. Fr{\"o}hlich,
\newblock Nucl. Phys. B {\bf 215 [FS7]} (1983) 209.

\bibitem{De92}
T.~A. DeGrand,
\newblock The present and future of lattice qcd,
\newblock in {\em Computational Methods in Field Theory Lecture Notes in
  Physics 409}, edited by H. Gausterer and C.~B. Lang, page 159,
  Springer-Verlag, Berlin, Heidelberg, 1992.

\bibitem{FaMaPa82}
M. Falcioni, E. Marinari, M.~L. Paciello, G. Parisi, and B. Taglienti,
\newblock Phys. Lett. B {\bf 108} (1982) 331.

\bibitem{FeSw88}
A.~M. Ferrenberg and R.~H. Swendsen,
\newblock Phys. Rev. Lett. {\bf 61} (1988) 2635.

\bibitem{FeSw89}
A.~M. Ferrenberg and R.~H. Swendsen,
\newblock Computers in Physics {\bf Sept./Oct.} (1989) 101.

\bibitem{FeSw89a}
A.~M. Ferrenberg and R.~H. Swendsen,
\newblock Phys. Rev. Lett. {\bf 63} (1989) 1195.

\bibitem{FiWo92}
H.~R. Fiebig and R.~M. Woloshyn,
\newblock Nucl. Phys. B (Proc. Suppl.) {\bf 30} (1993) 883.

\bibitem{FiWoDo92}
H.~R. Fiebig, R.~M. Woloshyn, and A. Dominguez,
\newblock Meson-meson scattering phase shifts in 2+1 dimensional lattice qed,
\newblock preprint FIU-PHY-92-10-23, 1992.

\bibitem{Fi68}
M.~E. Fisher,
\newblock in {\em Lectures in Theoretical Physics}, edited by W.~E. Brittin,
  volume VIIC, page~1, Gordon and Breach, New York, 1968.

\bibitem{Fi72}
M.~E. Fisher,
\newblock in {\em Critical Phenomena, Proc. of the 51th Enrico Fermi Summer
  School, Varena}, edited by M.~S. Green, Academic Press, New York, 1972.

\bibitem{FiRa86}
M.~E. Fisher and M. Randeria,
\newblock Phys. Rev. Lett. {\bf 56} (1986) 2332.

\bibitem{FrJaJe90}
C. Frick et~al.,
\newblock Nucl. Phys. B {\bf 331} (1990) 515.

\bibitem{Fr82}
J. Fr{\"o}hlich,
\newblock Nucl. Phys. B {\bf 200 [FS5]} (1982) 281.

\bibitem{GaLe84}
J. Gasser and H. Leutwyler,
\newblock Ann. Phys. {\bf 158} (1984) 142.

\bibitem{GaLe87}
J. Gasser and H. Leutwyler,
\newblock Phys. Lett. B {\bf 184} (1987) 83.

\bibitem{GaLe88}
J. Gasser and H. Leutwyler,
\newblock Nucl. Phys. B {\bf 307} (1988) 763.

\bibitem{GaHiLa93}
C.~R. Gattringer, I. Hip, and C.~B. Lang,
\newblock Nucl. Phys. B (Proc. Suppl.) {\bf 20} (1993) 875.

\bibitem{GaLa92}
C.~R. Gattringer and C.~B. Lang,
\newblock Phys. Lett. B {\bf 274} (1992) 95.

\bibitem{GaLa93}
C.~R. Gattringer and C.~B. Lang,
\newblock Nucl. Phys. B {\bf 391} (1993) 463.

\bibitem{GaLa87}
H. Gausterer and C.~B. Lang,
\newblock Phys. Lett. B {\bf 186} (1987) 103.

\bibitem{GaKu85b}
K. Gawedzki and A. Kupiainen,
\newblock Phys. Rev. Lett. {\bf 54} (1985) 92.

\bibitem{GlJa75}
J. Glimm and A. Jaffe,
\newblock Ann. Inst. H. Poincar{\'e} {\bf A22} (1975) 13.

\bibitem{GlJa76}
J. Glimm and A. Jaffe,
\newblock Fortschr. Phys. {\bf 21} (1976) 327.

\bibitem{GlJa87}
J. Glimm and A. Jaffe,
\newblock {\em Quantum Physics. A Functional Integral Point of View. 2nd
  Edition},
\newblock Springer-Verlag, New York, 1987.

\bibitem{Go90}
M. G{\"o}ckeler,
\newblock Nucl. Phys. B (Proc. Suppl.) {\bf 17} (1990) 347.

\bibitem{GoKaNe92}
M. G{\"o}ckeler, H. Kastrup, T. Neuhaus, and F. Zimmermann,
\newblock Nucl. Phys. B (Proc. Suppl.) {\bf 26} (1992) 516.

\bibitem{GoKaNe93}
M. G{\"o}ckeler, H.~A. Kastrup, T. Neuhaus, and F. Zimmermann,
\newblock Nucl. Phys. B {\bf 404} (1993) 517.

\bibitem{GoLe91}
M. G{\"o}ckeler and H. Leutwyler,
\newblock Nucl. Phys. B {\bf 350} (1991) 228.

\bibitem{GrPe78}
R.~B. Griffiths and P.~A. Pearce,
\newblock Phys. Rev. Lett. {\bf 41} (1978) 917.

\bibitem{GrPe79}
R.~B. Griffiths and P.~A. Pearce,
\newblock J. Stat. Phys. {\bf 20} (1979) 499.

\bibitem{GuPa87}
R. Gupta and A. Patel,
\newblock Phys. Lett. B {\bf 183} (1987) 193.

\bibitem{Ha87a}
T. Hara,
\newblock J. Stat. Phys. {\bf 47} (1987) 57.

\bibitem{HaTa87}
T. Hara and H. Tasaki,
\newblock J. Stat. Phys. {\bf 47} (1987) 99.

\bibitem{HaHa80}
A. Hasenfratz and P. Hasenfratz,
\newblock Phys. Lett. B {\bf 93} (1980) 165.

\bibitem{HaHa86}
A. Hasenfratz and P. Hasenfratz,
\newblock Nucl. Phys. B {\bf 270 [FS16]} (1986) 687.

\bibitem{HaHaHe84}
A. Hasenfratz, P. Hasenfratz, U. Heller, and F. Karsch,
\newblock Phys. Lett. B {\bf 140} (1984) 76.

\bibitem{HaJaJe91}
A. Hasenfratz et~al.,
\newblock Nucl. Phys. B {\bf 356} (1991) 332.

\bibitem{HaJaLa87}
A. Hasenfratz, K. Jansen, C.~B. Lang, T. Neuhaus, and H. Yoneyama,
\newblock Phys. Lett. B {\bf 199} (1987) 531.

\bibitem{Ha89}
P. Hasenfratz,
\newblock Nucl. Phys. B (Proc. Suppl.) {\bf 9} (1989) 3.

\bibitem{HaLe90}
P. Hasenfratz and H. Leutwyler,
\newblock Nucl. Phys. B {\bf 343} (1990) 241.

\bibitem{HaNa88}
P. Hasenfratz and J. Nager,
\newblock Z. Physik C {\bf 37} (1988) 477.

\bibitem{HaNi93}
P. Hasenfratz and F. Niedermayer,
\newblock Perfect lattice action for asymptotically free theories,
\newblock preprint BUTP-93/17, 1993.

\bibitem{He90}
U.~M. Heller,
\newblock Nucl. Phys. B (Proc. Suppl.) {\bf 17} (1990) 649.

\bibitem{He93}
U.~M. Heller,
\newblock Status of the higgs mass bound,
\newblock preprint FSU-SCRI-93-144, talk presented at LATTICE 93 in Dallas,
  1993.

\bibitem{HeKlNe92}
U.~M. Heller, M. Klomfass, H. Neuberger, and P. Vranas,
\newblock Nucl. Phys. B (Proc. Suppl.) {\bf 26} (1992) 522.

\bibitem{HeKlNe93a}
U.~M. Heller, M. Klomfass, H. Neuberger, and P. Vranas,
\newblock Nucl. Phys. B {\bf 405} (1993) 555.

\bibitem{HeNeVr92}
U.~M. Heller, H. Neuberger, and P. Vranas,
\newblock Phys. Lett. B {\bf 283} (1992) 335.

\bibitem{HiSc83}
J.~E. Hirsch and S.~H. Shenker,
\newblock Phys. Rev. B {\bf 27} (1983) 1736.

\bibitem{Hu82}
B. Hu,
\newblock Phys. Rep. {\bf 91C} (1982) 233.

\bibitem{ItDr89}
C. Itzykson and J.-M. Drouffe,
\newblock {\em Statistical Field Theory},
\newblock Cambridge University Press, Cambridge, 1989.

\bibitem{JaKuLi93a}
K. Jansen, J. Kuti, and C. Liu,
\newblock Phys. Lett. B {\bf 309} (1993) 119.

\bibitem{JaKuLi93}
K. Jansen, J. Kuti, and C. Liu,
\newblock Nucl. Phys. B (Proc. Suppl.) {\bf 30} (1993) 681.

\bibitem{JaLa91}
K. Jansen and C.~B. Lang,
\newblock Phys. Rev. Lett. {\bf 66} (1991) 3008.

\bibitem{FoKa69}
P.~W. Kasteleyn and C.~M. Fortuin,
\newblock J. Phys. Soc. Jpn. (Suppl.) {\bf 26} (1969) 11.

\bibitem{FoKa72}
P.~W. Kasteleyn and C.~M. Fortuin,
\newblock Physica (Utrecht) {\bf 57} (1972) 536.

\bibitem{Ke93}
R. Kenna,
\newblock Finite-size scaling in the {$O(N)$} {$\phi^4_4$} model,
\newblock hep-lat/9307003, 1993.

\bibitem{KeLa91}
R. Kenna and C.~B. Lang,
\newblock Phys. Lett. B {\bf 264} (1991) 396.

\bibitem{KeLa93a}
R. Kenna and C.~B. Lang,
\newblock Nucl. Phys. B (Proc. Suppl.) {\bf 30} (1993) 697.

\bibitem{KeLa93}
R. Kenna and C.~B. Lang,
\newblock Nucl. Phys. B {\bf 393} (1993) 461.

\bibitem{KuLiSh88}
J. Kuti, L. Lin, and Y. Shen,
\newblock Nucl. Phys. B (Proc. Suppl.) {\bf 4} (1988) 397.

\bibitem{KuLiSh88b}
J. Kuti, L. Lin, and Y. Shen,
\newblock Phys. Rev. Lett. {\bf 61} (1988) 678.

\bibitem{La85a}
C.~B. Lang,
\newblock Phys. Lett. B {\bf 155} (1985) 399.

\bibitem{La86c}
C.~B. Lang,
\newblock Nucl. Phys. B {\bf 265 [FS15]} (1986) 630.

\bibitem{La89}
C.~B. Lang,
\newblock Phys. Lett. B {\bf 229} (1989) 97.

\bibitem{LaSa88}
C.~B. Lang and M. Salmhofer,
\newblock Phys. Lett. B {\bf 205} (1988) 329.

\bibitem{LaWi93}
C.~B. Lang and U. Winkler,
\newblock Phys. Rev. D {\bf 47} (1993) 4705.

\bibitem{Le88}
I.-H. Lee,
\newblock Nucl. Phys. B (Proc. Suppl.) {\bf 4} (1988) 373.

\bibitem{Le87}
H. Leutwyler,
\newblock Phys. Lett. B {\bf 189} (1987) 197.

\bibitem{Lu86a}
M. L{\"u}scher,
\newblock Commun. Math. Phys. {\bf 105} (1986) 153.

\bibitem{Lu89}
M. L{\"u}scher,
\newblock in {\em Fields, Strings and Critical Phenomena, Les Houches, Session
  XLIX, 1988}, edited by E. Br{\'e}zin and J. Zinn-Justin, Elsevier,
  Amsterdam:, 1989.

\bibitem{Lu91a}
M. L{\"u}scher,
\newblock Nucl. Phys. B {\bf 364} (1991) 237.

\bibitem{Lu91}
M. L{\"u}scher,
\newblock Nucl. Phys. B {\bf 354} (1991) 531.

\bibitem{LuWe87}
M. L{\"u}scher and P. Weisz,
\newblock Nucl. Phys. B {\bf 290 [FS20]} (1987) 25.

\bibitem{LuWe88}
M. L{\"u}scher and P. Weisz,
\newblock Nucl. Phys. B {\bf 295 [FS21]} (1988) 65.

\bibitem{LuWe89}
M. L{\"u}scher and P. Weisz,
\newblock Nucl. Phys. B {\bf 318} (1989) 705.

\bibitem{LuWo90}
M. L{\"u}scher and U. Wolff,
\newblock Nucl. Pbys. B {\bf 339} (1990) 222.

\bibitem{Ma76a}
S.-K. Ma,
\newblock {\em Modern Theory of Critical Phenomena},
\newblock Benjamin/Cummings Publ. Co. Inc., Reading, MA, 1976.

\bibitem{Ma76}
S.-K. Ma,
\newblock Phys. Rev. Lett. {\bf 37} (1976) 461.

\bibitem{MaOl93}
F. Martinelli and E. Olivieri,
\newblock Some remarks on pathologies of renormalization-group transformations
  for the ising-model,
\newblock Univ. Roma preprint, 1993.

\bibitem{MeRoTe53}
N. Metropolis, A.~W. Rosenbluth, A.~H. Teller, and E. Teller,
\newblock J. Chem. Phys. {\bf 21} (1953) 1087.

\bibitem{MoWe87}
I. Montvay and P. Weisz,
\newblock Nucl. Phys. B {\bf 290 [FS20]} (1987) 327.

\bibitem{MuSh85}
G. Murthy and R. Shankar,
\newblock Phys. Rev. B {\bf 32} (1985) 5851.

\bibitem{Ne87}
H. Neuberger,
\newblock Phys. Lett. B {\bf 199} (1987) 536.

\bibitem{Ne88}
H. Neuberger,
\newblock Phys. Rev. Lett. {\bf 60} (1988) 889.

\bibitem{Ne90}
H. Neuberger,
\newblock Nucl. Phys. B (Proc. Suppl.) {\bf 17} (1990) 17.

\bibitem{Ne89b}
T. Neuhaus,
\newblock Nucl. Phys. B (Proc. Suppl.) {\bf 9} (1989) 21.

\bibitem{Ni92}
J. Nishimura,
\newblock Phys. Lett. B {\bf 294} (1992) 375.

\bibitem{OsSe78}
K. Osterwalder and E. Seiler,
\newblock Ann. Phys. {\bf 110} (1978) 440.

\bibitem{Re83}
C. Rebbi,
\newblock {\em Lattice Gauge Theories and Monte Carlo Simulations},
\newblock World Scientific, Singapore, 1983.

\bibitem{Ro91}
G. Roepstorff,
\newblock {\em Pfadintegrale in der Quantenphysik},
\newblock Vieweg: Braunschweig, 1991.

\bibitem{Ro92}
H.~J. Rothe,
\newblock {\em Lattice Gauge Theories - An Introduction},
\newblock World Scientific, Singapore, 1992.

\bibitem{SaMiJi77}
M. Sato, T. Miwa, and M. Jimbo,
\newblock Proc. Japan Acad. , Ser. A {\bf 53} (1977) 6.

\bibitem{Se82}
E. Seiler,
\newblock {\em Gauge Theories as a Problem of Constructive Quantum Field Theory
  and Statistical Mechanics, Springer Lecture Notes in Physics 159},
\newblock Springer, Berlin, 1982.

\bibitem{ShGuMu85}
R. Shankar, R. Gupta, and G. Murthy,
\newblock Phys. Rev. Lett. {\bf 55} (1985) 1812.

\bibitem{ShTo80}
S.~H. Shenker and J. Tobochnik,
\newblock Phys. Rev. B {\bf 22} (1980) 4462.

\bibitem{So89}
A. Sokal,
\newblock Monte carlo methods in statistical physics: Foundations and new
  algorithms,
\newblock Lecture Notes Cours de Troisieme Cycle de la Physique en Suisse
  Romande, Lausanne, 1989.

\bibitem{So91}
A. Sokal,
\newblock Nucl. Phys. B (Proc. Suppl.) {\bf 20} (1991) 55.

\bibitem{St71}
H. Stanley,
\newblock {\em Introduction to Phase Transitions and Critical Phenomena},
\newblock Oxford University Press, New York, 1971.

\bibitem{St79}
D. Stauffer,
\newblock Phys. Rep. {\bf 54} (1979) 1.

\bibitem{Sw83}
M. Sweeny,
\newblock Phys. Rev. B {\bf 27} (1983) 4445.

\bibitem{Sw79}
R.~H. Swendsen,
\newblock Phys. Rev. Lett. {\bf 42} (1979) 859.

\bibitem{Sw84}
R.~H. Swendsen,
\newblock Phys. Rev. Lett. {\bf 52} (1984) 1165.

\bibitem{Sw84a}
R.~H. Swendsen,
\newblock Phys. Rev. Lett. {\bf 52} (1984) 2321.

\bibitem{SwWa87}
R.~H. Swendsen and J.-S. Wang,
\newblock Phys. Rev. Lett. {\bf 58} (1987) 86.

\bibitem{Sy66}
K. Symanzik,
\newblock J. Math. Phys. {\bf 7} (1966) 510.

\bibitem{Sy69}
K. Symanzik,
\newblock in {\em Proc. Int. School of Physics ``Enrico Fermi'', Varenna Course
  XLV}, edited by R. Jost, Academic Press, 1969.

\bibitem{Sy83}
K. Symanzik,
\newblock Nucl. Phys. B {\bf 226} (1983) 187.

\bibitem{Ts85}
M.~M. Tsypin,
\newblock The effective potential of the lattice $\phi^4$ theory and the upper
  bound on the higgs mass,
\newblock Lebedev Phys. Inst. Moscow preprint 280, 1985.

\bibitem{EnFeSo91}
A.~C.~D. van Enter, R. Fern{\'a}ndez, and A.~D. Sokal,
\newblock Phys. Rev. Let. {\bf 66} (1991) 3253.

\bibitem{EnFeSo93}
A.~C.~D. van Enter, R. Fern{\'a}ndez, and A.~D. Sokal,
\newblock to appear in: J. Stat. Phys  (1993).

\bibitem{StWa90}
J.-S. Wang and D. Stauffer,
\newblock Z. Phys. B {\bf 78} (1990) 145.

\bibitem{WeZiGo93}
J. Westphalen, F. Zimmermann, M. G{\"o}ckeler, and H.~A. Kastrup,
\newblock Scattering phases in the broken phase of the 4-d o(4) non-linear
  $\sigma$-model,
\newblock Contrib. to LATTICE 93 in Dallas, 1993.

\bibitem{Wh84}
C. Whitner,
\newblock PhD Thesis, Princeton University, 1984.

\bibitem{Wi71}
K.~G. Wilson,
\newblock Phys. Rev. B {\bf 4} (1971) 3184.

\bibitem{Wi74}
K.~G. Wilson,
\newblock Phys. Rev. D {\bf 10} (1974) 2445.

\bibitem{WiKo74}
K.~G. Wilson and J. Kogut,
\newblock Phys. Rep. {\bf 12C} (1974) 76.

\bibitem{Wo89}
U. Wolff,
\newblock Phys. Rev. Lett. {\bf 62} (1989) 361.

\bibitem{Wo92}
U. Wolff,
\newblock High precision simulations with fast algorithms,
\newblock in {\em Computational Methods in Field Theory, Lecture Notes in
  Physics 409}, edited by H. Gausterer and C.~B. Lang, page 127,
  Springer-Verlag, Berlin, Heidelberg, 1992.

\bibitem{YaLe52}
C.~N. Yang and T.~D. Lee,
\newblock Phys. Rev. {\bf 87} (1952) 404.

\bibitem{YaLe52a}
C.~N. Yang and T.~D. Lee,
\newblock Phys. Rev. {\bf 87} (1952) 410.

\bibitem{Zi93}
F. Zimmermann,
\newblock Unstable particles in finite volume: the broken phase of the 4-d o(4)
  non-linear $\sigma$-model,
\newblock RHTH Aachen PhD Thesis, 1993.

\bibitem{ZiWeGo93}
F. Zimmermann, J. Westphalen, M. G{\"o}ckeler, and H.~A. Kastrup,
\newblock Nucl. Phys. B (Proc. Suppl.) {\bf 30} (1993) 879.

\end{thebibliography}
\end{document}